\documentclass[a4paper,12pt,reqno]{article}
\usepackage{amsmath,amsfonts,amssymb,amsthm,dsfont}

\usepackage{hyperref}
\allowdisplaybreaks
\numberwithin{equation}{section}

\newcommand{\om}{\omega}
\newcommand{\half}{\tfrac{1}{2}}
\newcommand{\ihalf}{\tfrac{\mathrm{i}}{2}}
\newcommand{\Ii}{\mathrm{i}}

\newcommand{\6}{\partial}
\newcommand{\7}{\hat}

\newcommand{\ua}{{\underline{\alpha}}}
\newcommand{\ub}{{\underline{\beta}}}
\newcommand{\ug}{{\underline{\gamma}}}
\newcommand{\ud}{{\underline{\delta}}}
\newcommand{\ul}[1]{{\underline{#1}}}
\newcommand{\ol}[1]{{\overline{#1}}}
\newcommand{\com}[2]{[\,#1\, ,\,#2\,]}	% commutator
\newcommand{\acom}[2]{\{#1\, ,\,#2\}}	% anticommutator

\newcommand{\gam}{\Gamma}	% gamma-matrices
\newcommand{\CC}{C}	% charge conjugation matrix
\newcommand{\IC}{C^{-1}}	% inverse charge conjugation matrix
\newcommand{\unit}{\mathds{1}}

\newcommand{\Dim}{D} % spacetime dimension
\newcommand{\Ns}{N} % number of susy sets
\newcommand{\Omg}{\Omega_\mathrm{gh}} % space of ghost polynomials
\newcommand{\fb}{s_\mathrm{gh}} % coboundary operator
\newcommand{\Hg}{H_\mathrm{gh}} % cohomology in the space of ghost polynomials
\newcommand{\hHg}{\hat H_\mathrm{gh}} % cohomology in the space of ghost polynomials
\newcommand{\cdeg}{$c$-degree} % degree in translation ghosts
\newcommand{\LRA}{\Leftrightarrow}
\newcommand{\Then}{\Rightarrow}
\newcommand{\Mod}{\ \mathrm{mod}\ }
\newcommand{\so}[1]{$\mathfrak{so}(#1)$}
\newcommand{\hOmg}{\hat\Omega}%{\hat\Omega_\mathrm{gh}} % space of ghost polynomials
\newcommand{\PF}[1]{{\bf Proof sketch for lemma #1:} }

\newcommand{\QED}{\hfill$\blacksquare$}

\newtheorem{prop}{Proposition}[section]
\newtheorem{lemma}[prop]{Lemma}

\newcommand{\THE}[2]{\Theta_{#1}^{(#2)}}
\newcommand{\hTHE}[2]{\hat\Theta{}_{#1}^{(#2)}}
\newcommand{\cTHE}[3]{\Theta{}_{#1}^{#2(#3)}}
\newcommand{\vTHE}[2]{\vartheta_{#1}^{#2}}

\begin{document}

%v2: 28 pages, comments and explanations added (at the end of sections 4, 6.1, 8.1, 9, 10.1, 11, beginning of section 5, refs. [11,12]), minor improvements of text and formulas

\begin{flushright}
ITP--UH--08/13
\end{flushright}

\begin{center}
 {\large\bfseries Supersymmetry algebra cohomology IV:\\[6pt] Primitive elements in all dimensions from $\Dim=4$ to $\Dim=11$}
 \\[5mm]
 Friedemann Brandt \\[2mm]
 \textit{Institut f\"ur Theoretische Physik, Leibniz Universit\"at Hannover, Appelstra\ss e 2, D-30167 Hannover, Germany}
\end{center}

\begin{abstract}
The primitive elements of the supersymmetry algebra cohomology as defined in previous work are derived for standard supersymmetry algebras in dimensions $\Dim=5,\dots,11$ for all signatures of the related Clifford algebras of gamma matrices and all numbers of supersymmetries. The results are presented in a uniform notation along with results of previous work for $\Dim=4$, and derived by means of dimensional extension from $\Dim=4$ up to $\Dim=11$. 
\end{abstract}

%\newpage
\setcounter{tocdepth}{1}
\tableofcontents

\section{Introduction}

This paper relates to supersymmetry algebra cohomology \cite{Brandt:2009xv} for supersymmetry algebras in dimensions $\Dim=4,\dots,11$ of translational generators $P_a$ ($a=1,\dots,\Dim)$ and supersymmetry generators $Q^i_\ua$ ($\ua=\ul 1,\dots,\ul{2^{\lfloor \Dim/2 \rfloor}}$; $i=1,\dots,\Ns$) of the form
 \begin{align}
  \com{P_a}{P_b}=0,\quad \com{P_a}{Q^i_\ua}=0,\quad \acom{Q^i_\ua}{Q^j_\ub}=M^{ij}\,(\gam^a \IC)_{\ua\ub}P_a
	\label{i4-1} 
 \end{align}
where $M^{ij}$ are the entries of an $\Ns\times\Ns$ matrix $M$ given by
  \begin{align}
  \Dim\in\{4,8,9,10,11\}:\ M=-\Ii\, \unit_{\Ns\times\Ns},\quad
  \Dim\in\{5,6,7\}:\ M=\unit_{\Ns/2\times\Ns/2}\otimes\sigma_2     
	\label{i4-2} 
 \end{align}
with $\unit_{n\times n}$ denoting the $n\times n$ unit matrix, $\sigma_2$ denoting the second Pauli-matrix, see \eqref{sigmas}, and $\Ii$ denoting the imaginary unit. The difference between $M$ in $\Dim=5,6,7$ and in $\Dim=4,8,9,10,11$ originates from the symmetry properties of the matrices $\gam^a \IC$: in $\Dim=5,6,7$ these matrices are antisymmetric whereas in $\Dim=4,8,9,10,11$ they are symmetric  (in $\Dim=4$ and $\Dim=8$ this is due to our choice of $\CC$ \cite{note0}). 

The object of this paper is the determination of the primitive elements of the supersymmetry algebra cohomology for the supersymmetry algebras \eqref{i4-1} under study, for all signatures $(t,\Dim-t)$ ($t=0,\dots,\Dim $) of the Clifford algebra of the $\gam^a$ and for all respectively possible values of $\Ns$.  
According to our definition \cite{Brandt:2009xv} these primitive elements represent the cohomology $\Hg(\fb)$ of the coboundary operator
 \begin{align}
  \fb=-\half M^{ij}\,(\gam^a \IC)_{\ua\ub}\,\xi^\ua_i\xi^\ub_j\,\frac{\6}{\6 c^a}
	\label{i4-4} 
 \end{align}
in the space $\Omg$ of polynomials (with coefficients in $\mathbb{C}$) in anticommuting translation ghosts $c^a$ and commuting supersymmetry ghosts $\xi^\ua_i$ corresponding to the translational generators $P_a$ and supersymmetry generators $Q^i_\ua$, respectively.

Depending on the dimension $\Dim$ and signature $(t,\Dim-t)$, the $Q^i$ and $\xi_i$ are Majorana Weyl (MW), symplectic Majorana Weyl (SMW), Majorana (M) or symplectic Majorana (SM) supersymmetries according to table \eqref{i4-3} which also contains the coefficients $\eta$ and $\epsilon$ related to the charge conjugation matrix $\CC$ ($\gam_a^{\top}=-\eta\,\CC\,\gam_a\IC$, $\CC^{\top}=-\epsilon\, \CC$ \cite{Brandt:2009xv,VanProeyen:1999ni}):
 \begin{align}
 \begin{array}{|c|c|c|l|l|l|l|}
 \hline
 \Dim&\eta & \epsilon  & t\Mod 4=0 &t\Mod 4=1 &t\Mod 4=2 &t\Mod 4=3\\
 \hline
 4 & +1 & +1 &  \mathrm{SM}_4(2\mathbb{N}) & \mathrm{M}_4(\mathbb{N}) & \mathrm{M}_4(\mathbb{N})& \mathrm{M}_4(\mathbb{N})\\
 \hline
 5 & -1 & +1 &  \mathrm{SM}_4(2\mathbb{N}) & \mathrm{SM}_4(2\mathbb{N}) & \mathrm{M}_4(2\mathbb{N})& \mathrm{M}_4(2\mathbb{N})\\
 \hline
 6 & +1 & -1 &  \mathrm{M}_8(2\mathbb{N}) & \mathrm{SMW}_4(2\mathbb{N}) & \mathrm{M}_8(2\mathbb{N})& \mathrm{MW}_4(2\mathbb{N})\\
 \hline
 7 & +1 & -1 &  \mathrm{M}_8(2\mathbb{N}) & \mathrm{SM}_8(2\mathbb{N}) & \mathrm{SM}_8(2\mathbb{N})& \mathrm{M}_8(2\mathbb{N})\\
 \hline
 8 & -1 & -1 & \mathrm{M}_{16}(\mathbb{N}) & \mathrm{M}_{16}(\mathbb{N})& \mathrm{SM}_{16}(2\mathbb{N})& \mathrm{M}_{16}(\mathbb{N})\\
 \hline
 9 & -1 & -1 & \mathrm{M}_{16}(\mathbb{N}) & \mathrm{M}_{16}(\mathbb{N})& \mathrm{SM}_{16}(2\mathbb{N})& \mathrm{SM}_{16}(2\mathbb{N})\\
 \hline
 10 & +1 & +1 & \mathrm{M}_{32}(\mathbb{N}) & \mathrm{MW}_{16}(\mathbb{N}) & \mathrm{M}_{32}(\mathbb{N}) & \mathrm{SMW}_{16}(2\mathbb{N})\\
 \hline
 11 & +1 & +1 & \mathrm{SM}_{32}(2\mathbb{N}) & \mathrm{M}_{32}(\mathbb{N}) & \mathrm{M}_{32}(\mathbb{N}) & \mathrm{SM}_{32}(2\mathbb{N})\\
 \hline
 \end{array}
	\label{i4-3} 
 \end{align}
In \eqref{i4-3} the subscripts denote the number of independent spinor components of the respective spinors, and $\mathbb{N}=\{1,2,\ldots\}$ or $2\mathbb{N}=\{2,4,\ldots\}$ in parantheses indicate the possible values of $\Ns$ for the particular signature $(t,\Dim-t)$. For instance, $\mathrm{SM}_4(2\mathbb{N})$ for signature $(0,4)$ in $\Dim=4$ indicates that in this case the $Q^i$ and $\xi_i$ are symplectic Majorana spinors, each $Q^i$ and $\xi_i$ has four independent spinor components, and $\Ns\in\{2,4,\ldots\}$, i.e. $\Ns$ can take the values $2, 4, \dots$

The use of Majorana or symplectic Majorana spinors has the advantage that we need not worry about the complex conjugated spinors $Q^*$, $\xi^*$ because their components are related one to one to the components of the $Q$, $\xi$ by the Majorana or symplectic Majorana condition, i.e. we can use the components of the $Q$, $\xi$ as a complete set of independent components of the supersymmetries and supersymmetry ghosts, respectively.

In $\Dim=4$ and $\Dim=8$ we use, for all signatures, Majorana or symplectic Majorana spinors consisting of two Weyl spinors with opposite chiralities, even when there are Majorana Weyl or symplectic Majorana Weyl spinors (which is the case for $t=0,2,\dots$). The reason is that in $\Dim=4,8,\dots$ the matrices $\gam^a \IC$ in \eqref{i4-1} relate Weyl spinors of opposite chiralities (see $p=1$ in \eqref{dt3}). Therefore, in $\Dim=4$ and $\Dim=8$ any nontrivial anticommutator in an algebra \eqref{i4-1} relates two Weyl supersymmetries with opposite chiralities which can be combined to one Majorana or symplectic Majorana supersymmetry.  

In contrast, in $\Dim=2,6,\dots$ the matrices $\gam^a \IC$ relate supersymmetries with equal chirality. Therefore, in $\Dim=6$ and $\Dim=10$ for signatures with $t=1,3,\dots$ one cannot always combine the Majorana Weyl or symplectic Majorana Weyl supersymmetries to a set of ordinary Majorana or symplectic Majorana supersymmetries containing two Weyl spinors with opposite chiralities (this can only be achieved when there are equal numbers of Majorana Weyl or symplectic Majorana Weyl supersymmetries with opposite chiralities). Hence, in $\Dim=6$ and $\Dim=10$ we treat the signatures with $t=1,3,\dots$ differently from those with $t=0,2,\dots$ by using Majorana Weyl or symplectic Majorana Weyl spinors in the former cases and ordinary Majorana or symplectic Majorana spinors in the latter cases, see table \eqref{i4-3}. As a consequence, in $\Dim=6$ and $\Dim=10$ the value of $\Ns$ by itself does not determine the number of independent spinor components of supersymmetries because this number also depends on the signature. 

$\Hg(\fb)$ is computed in $\Dim=5,\dots,11$ by ``dimensional extension'' (termed ``dimension climbing'' in ref. \cite{Brandt:2010fa}): we relate the cohomology in $\Dim$ dimensions to the cohomology in $\Dim-1$ dimensions  and use the results in $\Dim-1$ dimensions to derive the results in $\Dim$ dimensions. This method is outlined in section \ref{climb}.

Section \ref{defs} compiles definitions and facts which are used later on. 
In sections \ref{4D} and \ref{5D} results for $\Dim=4$ and $\Dim=5$ derived in ref. \cite{Brandt:2010tz} are reformulated in terms of objects introduced in section \ref{defs}, and completed for $\Dim=5$. Sections \ref{6D} to \ref{11D} present the results for dimensions $\Dim=6,\dots,11$. The derivation of these results is outlined in a comprehensible but condensed form in order to keep the paper reasonably short and readable (the presentation for $\Dim=6$ in section \ref{6D} is more detailed in order to exemplify the derivation of the results for one case).

Some of the results derived here for $\Dim\geq 6$ were found already in refs.
\cite{Movshev:2010mf,Movshev:2011pr}. These results concern the lowest possible number of supersymmetries in each of these dimensions and are thus for $\Dim\neq 7$ limited to particular signatures $(t,\Dim-t)$, see \eqref{i4-3}. For instance, the results presented in refs. \cite{Movshev:2010mf,Movshev:2011pr} apply in $\Dim=6$ to the cases $\Ns=2$ for signatures $(1,5)$, $(3,3)$, $(5,1)$, and in $\Dim=10$ to the cases $\Ns=1$ for signatures $(1,9)$, $(5,5)$, $(9,1)$ but not to other values of $\Ns$ or other signatures, respectively. The results derived in the present work confirm and specify the results for $\Dim=6,7,9,10,11$ presented in refs. \cite{Movshev:2010mf,Movshev:2011pr}, and extend them to all numbers of supersymmetries and all signatures. The results for $\Dim=8$ are commented on at the end of section \ref{8D}.

The results are particularly useful in the context of BRST cohomological analyses of globally and locally supersymmetric field theories \cite{Brandt:2009xv,Brandt:2012np}.

\section{\texorpdfstring{Dimensional extension from $\Dim=4$ up to $\Dim=11$}{Dimensional extension from D=4 up to D=11}}\label{climb}

We shall now explain our method to compute $\Hg(\fb)$ in $\Dim=5,\dots,11$ dimensions by means of the results in $\Dim-1$ dimensions, respectively.
We study the cocycle condition in $\Hg(\fb)$ separately for each \cdeg\ (= degree in the translation ghosts) which is possible since $\fb$ decreases the \cdeg\ by one unit. The \cdeg\ is denoted by a superscript. The subspace of $\Omg$ containing the polynomials of \cdeg\ $p$ is denoted by $\Omg^p$. $\Hg^p(\fb)$ denotes the cohomology of $\fb$ in $\Omg^p$ \cite{note0a}. The cocycle conditions to be studied thus read
\begin{align}
  \fb\om^p=0,\ \ \om^p\in\Omg^p=\{\om\in\Omg\,|\ N_c\, \om=p\,\om\}
	\label{c1} 
\end{align}
where $N_c=c^a\frac{\6}{\6 c^a}$ denotes the counting operator for the translation ghosts.

In order to relate $\Hg(\fb)$ in $\Dim$ and $\Dim-1$ dimensions
we define the subspace $\hOmg$ of ghost polynomials in $\Dim$ dimensions which do not depend on the translation ghost $c^\Dim$, as well as the subspaces $\hOmg^p\subset\Omg^p$ thereof, 
\begin{align}
  \hOmg=\Big\{\om\in\Omg\,\Big|\ \frac{\6\om}{\6 c^\Dim}=0\Big\},\
  \hOmg^p=\{\om\in\hOmg\,|\ N_c\, \om=p\,\om\}.
	\label{c3} 
\end{align}
As a ghost polynomial in $\Omg^p$ is at most linear in $c^\Dim$, it can be uniquely written as
\begin{align}
  \om^p=c^\Dim\7\om^{p-1}+\7\om^p,\quad \7\om^{p-1}\in\hOmg^{p-1},\ \7\om^p\in\hOmg^p.
	\label{c4} 
\end{align}
This gives:
\begin{align}
  \fb\om^p=(\fb c^\Dim)\7\om^{p-1}-c^\Dim(\fb\7\om^{p-1})+\fb\7\om^p.
	\label{c5} 
\end{align}
As $\fb c^\Dim$ is a quadratic polynomial in the $\xi_i^\ua$,
only the second term on the right hand side of \eqref{c5} contains $c^\Dim$. Hence, the cocycle condition \eqref{c1} splits into two conditions:
\begin{align}
  &\fb\7\om^{p-1}=0,\label{c6}\\
  &(\fb c^\Dim)\7\om^{p-1}+\fb\7\om^p=0.\label{c7} 
\end{align}
\eqref{c6} and \eqref{c7} relate $\Hg(\fb)$ to the cohomology of $\fb$ in $\hOmg$ which we denote by $\hHg(\fb)$. Accordingly the cohomology of $\fb$ in $\hOmg^p$ is denoted by $\hHg^p(\fb)$.
By \eqref{c6} the constituent $\7\om^{p-1}$ of $\om^p$ is a cocycle in $\hHg^{p-1}(\fb)$. Furthermore, any contribution $\fb\7\eta^p$ to $\7\om^{p-1}$ with $\7\eta^p\in\hOmg^p$ can be removed from $\om^p$ by adding the coboundary $\fb(c^\Dim\7\eta^p)$ owing to $\om^p+\fb(c^\Dim\7\eta^p)=c^\Dim(\7\om^{p-1}-\fb\7\eta^p)+\7\om^{p\,\prime}$, with $\7\om^{p\,\prime}=\7\om^p+(\fb c^\Dim)\7\eta^p\in\hOmg^p$ redefining the constituent $\7\om^p$ of $\om^p$. Hence, the constituent $\7\om^{p-1}$ of $\om^p$ can be assumed to be a nontrivial representative of $\hHg^{p-1}(\fb)$,
\begin{align}
  \7\om^{p-1}\in\hHg^{p-1}(\fb).\label{c8} 
\end{align}
\eqref{c7} imposes that $(\fb c^\Dim)\7\om^{p-1}$ is trivial in $\hHg^{p-1}(\fb)$ which in general can impose an extra condition on $\7\om^{p-1}$ (in addition to \eqref{c8}). Furthermore, for any $(\fb c^\Dim)\7\om^{p-1}$ which is trivial in $\hHg^{p-1}(\fb)$ we may consider \eqref{c7} as an inhomogeneous equation for $\7\om^p$ whose solution is the sum of the general solution $\7\om^p_\mathrm{hom}$ of the homogeneous equation $\fb\7\om^p_\mathrm{hom}=0$ and a particular solution $\7\om^p_\mathrm{part}$ of \eqref{c7}. Moreover, any contribution $\fb\7\eta^{p+1}$ to $\7\om^p_\mathrm{hom}$ with $\7\eta^{p+1}\in\hOmg^{p+1}$ can be removed from $\om^p$ by subtracting the coboundary $\fb \7\eta^{p+1}$. Hence, $\7\om^p_\mathrm{hom}$ can be assumed to be a nontrivial representative of $\hHg^p(\fb)$. This gives
\begin{align}
  &\7\om^p=\7\om^p_\mathrm{hom}+\7\om^p_\mathrm{part}\,,\label{c9}\\ 
  &\7\om^p_\mathrm{hom}\in\hHg^p(\fb),\label{c10} \\
  &(\fb c^\Dim)\7\om^{p-1}+\fb\7\om^p_\mathrm{part}=0.\label{c11}
\end{align}
Equations \eqref{c6}--\eqref{c11} trace $\Hg(\fb)$ back to $\hHg(\fb)$. The crucial point is that $\hHg(\fb)$ in $\Dim\in\{5,\dots,11\}$ can be obtained from $\Hg(\fb)$ in $\Dim-1$ dimensions. This can be shown, for instance, by choosing $\gam$-matrices and $\CC$ in $\Dim$ dimensions according to equations \eqref{c12}--\eqref{c13c} which are compatible with \eqref{i4-3}:
\begin{align}
\Dim\Mod 2=1:\ &\gam^a=\gam^a_{(\Dim-1)}\ \mathrm{for}\ a\in\{1,\dots,\Dim-1\},\
                       \gam^\Dim=(k_\Dim)^{-1}\,\hat\gam_{(\Dim-1)}\label{c12}\\
\Dim\Mod 2=0:\ &\gam^a=\sigma_1\otimes\gam^a_{(\Dim-1)}\ \mathrm{for}\ a\in\{1,\dots,\Dim-1\},\notag\\
                   &\gam^\Dim=(k_\Dim)^{-1}\,\sigma_2\otimes\unit,\    
                     \hat\gam=\sigma_3\otimes\unit\label{c13}\\
\Dim\Mod 8\in\{1,3,7\}:\ &\CC=\CC_{(\Dim-1)}\label{c12a}\\
\Dim\Mod 8=5:\ &\CC=\CC_{(\Dim-1)}\hat\gam_{(\Dim-1)}\label{c12b}\\
\Dim\Mod 8=0:\ &\CC=\sigma_3\otimes\CC_{(\Dim-1)}\label{c13a}\\
\Dim\Mod 8\in\{2,6\}:\ &\CC=\Ii\,\sigma_2\otimes\CC_{(\Dim-1)}\label{c13b}\\
\Dim\Mod 8=4:\ &\CC=\sigma_0\otimes\CC_{(\Dim-1)}\label{c13c}
\end{align}
where
$\gam^a_{(\Dim-1)}$, $\hat\gam_{(\Dim-1)}$ and $\CC_{(\Dim-1)}$ denote the $\gam$-matrices and the charge conjugation matrix in $\Dim-1$ dimensions, in \eqref{c13}
$\unit$ denotes the $2^{\Dim/2-1}\times 2^{\Dim/2-1}$ unit matrix, 
 \begin{align}
  k_\Dim=\left\{\begin{array}{cl}
  \Ii & \mathrm{for}\ t=\Dim\\
  1& \mathrm{for}\ t<\Dim\end{array}\right.
  \label{c14}
 \end{align}
and
 \begin{align}
  \sigma_0=\begin{pmatrix} 1 & 0 \\ 0 & 1 \end{pmatrix} ,\quad
  \sigma_1=\begin{pmatrix} 0 & 1 \\ 1 & 0 \end{pmatrix} ,\quad
  \sigma_2=\begin{pmatrix} 0 & -\Ii \\ \Ii & 0 \end{pmatrix} ,\quad
  \sigma_3=\begin{pmatrix} 1 & 0 \\ 0 & -1 \end{pmatrix} .
	\label{sigmas} 
 \end{align}
Using \eqref{c12}--\eqref{c13c} and suitably relating the supersymmetry ghosts $\xi_i$ in $\Dim\in\{5,\dots,11\}$ to corresponding supersymmetry ghosts $\xi_{i(\Dim-1)}$ in $\Dim-1$ dimensions for appropriate values of $\Ns$, the action of $\fb$ in $\hOmg$ in $\Dim$ dimensions becomes identical to the action of $\fb$ in $\Omg$ in $\Dim-1$ dimensions, see sections 
\ref{5D}--\ref{11D} for details. This allows one to derive $\hHg(\fb)$ in $\Dim\in\{5,\dots,11\}$ from $\Hg(\fb)$ in $\Dim-1$ dimensions.

The decomposition \eqref{c4} is also useful for analysing the coboundary condition in $\Hg(\fb)$ in order to sieve the nontrivial cocycles. To this end the coboundary condition in dimension $\Dim$ at \cdeg\ $p$ is written as
 \begin{align}
  \om^p=\fb\eta^{p+1},\quad \eta^{p+1}=c^\Dim\7\eta^p+\7\eta^{p+1},\ 
  \7\eta^{p}\in\hOmg^{p},\ \7\eta^{p+1}\in\hOmg^{p+1}.
  \label{c15}
 \end{align}
Using \eqref{c4} this yields
 \begin{align}
 \7\om^{p-1}&=-\fb\7\eta^p,\label{c16}\\
 \7\om^p&=(\fb c^\Dim)\,\7\eta^p+\fb\7\eta^{p+1}.
  \label{c17}
 \end{align}
In particular this implies that a cocycle $\om^p$ in $\Hg(\fb)$ is nontrivial if its constituent $\7\om^{p-1}$ is nontrivial in $\hHg(\fb)$, and that $\7\om^p_\mathrm{hom}$ 
can be neglected if in $\hHg(\fb)$ it is equivalent to $(\fb c^\Dim)\,\7\eta^p_\mathrm{hom}$ for a cocycle $\7\eta^p_\mathrm{hom}$ in $\hHg(\fb)$.

{\em Comments:} \\
1. Suppose that $\hHg^p(\fb)$ vanishes for all $p\geq p_0$. Owing to \eqref{c6}--\eqref{c11} this implies that $\Hg^p(\fb)$ vanishes for all $p> p_0$:
 \begin{align}
  \forall\,p\geq p_0:\ \hHg^p(\fb)=0\quad \Then\quad \forall\,p> p_0:\ \Hg^p(\fb)=0.
  \label{c18}
 \end{align}

2. The case $p=1$ is somewhat special because in this case \eqref{c6} is automatically fulfilled since $\hat\om^0$ does not depend on translation ghosts at all. In other words, \eqref{c6} does not impose restrictions on the cocycles $\om^1$ at all, i.e. these cocycles have to be determined solely from \eqref{c7} modulo coboundaries.

3. The dimensional extension method outlined above  is applicable modulo 8 in the dimensions, i.e. it applies analogously to any sequence of dimensions $\Dim=4+8k,\dots,11+8k$.

4. Albeit we shall use spinor representations fulfilling \eqref{c12}--\eqref{c13c} to derive the results, these results extend to other spinor representations owing to the \so{t,\Dim-t}-covariance of the results.

5. The results for $\Hg(\fb)$ in $\Dim=4$ cannot be derived from the results in $\Dim=3$ by dimensional extension as outlined above. The reason is that in $\Dim=4$ the matrices $\gam^a\IC$ relate Weyl spinors of opposite chiralities and thus different 2-component spinors, whereas in $\Dim=3$ they relate equal 2-component spinors. For this reason $\Hg(\fb)$ in $\Dim=4$ was computed ``from scratch'' in ref. \cite{Brandt:2010tz}.

\section{Definitions and useful facts}\label{defs}

Unless specified otherwise, we use notation and conventions as in ref. \cite{Brandt:2009xv}. As in refs. \cite{Brandt:2010fa,Brandt:2010tz} $\sim$ denotes equivalence in $\Hg(\fb)$, i.e. for $\om_1,\om_2\in\Omg$ the notation $\om_1\sim \om_2$ means $\om_1-\om_2=\fb\om_3$ for some $\om_3\in\Omg$:
\begin{align}
  \om_1\sim \om_2\quad :\Leftrightarrow \quad \exists\,\om_3:\ \om_1-\om_2=\fb\om_3\quad (\om_1,\om_2,\om_3\in\Omg).
  \label{equiv}
\end{align}

In all dimensions we define the following \so{t,\Dim-t}-covariant ghost polynomials:
\begin{align}
  \vTHE{i}{}&
  =c^a\,\xi_i\,\gam_a\label{def1}\\
  \THE{ij}{p}&
  =\tfrac{1}{p!}\,c^{a_1} \ldots c^{a_p}\, \xi_i\,\gam_{a_1\ldots a_p}\IC\,\xi_j^\top\label{def2}\\
  \THE{ij,\,a_1\ldots a_k}{p}&
  =\tfrac{1}{(p-k)!}\,c^{a_{k+1}} \ldots c^{a_p}\, \xi_i\,\gam_{a_1\ldots a_p}\IC\,\xi_j^\top
  =\frac{\6}{\6 c^{a_k}}\ldots \frac{\6}{\6 c^{a_1}}\, \THE{ij}{p}\label{def3}
\end{align}
where $\gam_{a_1\ldots a_p}$ denotes the totally antisymmetrized product of $p$ gamma matrices
\begin{align} 
\gam_{a_1\ldots a_p}
=\gam_{[a_1}\ldots\gam_{a_p]}=\frac{1}{p!}\sum_{\sigma\in S_p}(-1)^{\mathrm{sgn}(\sigma)}\gam_{a_{\sigma(1)}}\ldots\gam_{a_{\sigma(p)}}
\label{def3a}
\end{align}
and we use matrix notation with $\xi_i$ denoting a ``row spinor'' and $\xi_i^\top$ a ``column spinor'',
\begin{align} 
\vTHE{i}{\ua}=c^a\,\xi_i^\ub\,\gam_{a\,\ub}{}^\ua\,,\quad
\xi_i\,\gam_{a_1\ldots a_p}\IC\,\xi_j^\top=\xi_i^\ua\,(\gam_{a_1\ldots a_p}\IC)_{\ua\ub}\,\xi_j^\ub\,.
\end{align}

In even dimensions we further define:
\begin{align}
  \xi^+_i&=\half\,\xi_i\,(\unit+\hat\gam)\label{def4}\\
  \xi^-_i&=\half\,\xi_i\,(\unit-\hat\gam)\label{def5}\\
  \vTHE{i}{+}&=\half\,\vTHE{i}{}\,(\unit+\hat\gam)
  =c^a\,\xi^-_i\,\gam_a\label{def6}\\
  \vTHE{i}{-}&=\half\,\vTHE{i}{}\,(\unit-\hat\gam)
  =c^a\,\xi^+_i\,\gam_a\label{def7}\\
  \hTHE{ij}{p}&
  =\tfrac{1}{p!}\,c^{a_1} \ldots c^{a_p}\, \xi_i\,\hat\gam\gam_{a_1\ldots a_p}\IC\,\xi_j^\top\label{def8}\\
  \hTHE{ij,\,a_1\ldots a_k}{p}&
  =\tfrac{1}{(p-k)!}\,c^{a_{k+1}} \ldots c^{a_p}\, \xi_i\,\hat\gam\gam_{a_1\ldots a_p}\IC\,\xi_j^\top
  =\frac{\6}{\6 c^{a_k}}\ldots \frac{\6}{\6 c^{a_1}}\, \hTHE{ij}{p}\label{def9}\\
  \cTHE{ij}+p&
  =\half (\THE{ij}p+\hTHE{ij}{p})\label{def10}\\  
  \cTHE{ij,\,a_1\ldots a_k}+{p}&
  =\half (\THE{ij,\,a_1\ldots a_k}{p}+\hTHE{ij,\,a_1\ldots a_k}{p})\label{def11}\\
  \cTHE{ij}-{p}&
  =\half (\THE{ij}{p}-\hTHE{ij}{p})\label{def12}\\
  \cTHE{ij,\,a_1\ldots a_k}-{p}&
  =\half (\THE{ij,\,a_1\ldots a_k}{p}-\hTHE{ij,\,a_1\ldots a_k}{p})\label{def13}
\end{align}

$\fb$ anticommutes with the derivatives with respect to the translation ghosts $c^a$. This implies that the first and higher order derivatives of a cocycle in $\Hg(\fb)$ with respect to the $c^a$ are also cocycles,
 \begin{align}
  \fb\om^p=0\quad\Then\quad  
  \forall k\in\{1,\dots,p\}:\ \fb\,\frac{\6^k\om^p}{\6 c^{a_k}\dots\6 c^{a_1}}=0.
  \label{c19}
 \end{align}
In particular, owing to \eqref{c19}, all $\THE{ij,\,a_1\ldots a_k}{p}$ for $k=1,\dots,p-1$ are cocycles in $\Hg(\fb)$ if $\THE{ij}{p}$ is a cocycle in $\Hg(\fb)$ and an analogous statement applies to the $\hTHE{ij,\,a_1\ldots a_k}{p}$.

For the reader's convenience we list some properties of the matrices $\gam_{a_1\ldots a_p}\IC$ and $\hat\gam\gam_{a_1\ldots a_p}\IC$ (valid for $\eta,\epsilon$ as in \eqref{i4-3}) which are useful for understanding properties of the above ghost polynomials depending on $\Dim$ and $p$:
 
(i) The $\gam_{a_1\ldots a_p}\IC$ are symmetric ($S$) or antisymmetric ($A$) in their spinor indices:  
 \begin{align}
 \begin{array}{|l|c|c|c|c|}
 \hline
 &p\Mod 4=0&p\Mod 4=1&p\Mod 4=2&p\Mod 4=3\\
 \hline
 \Dim=4&A&S&S&A\\
 \hline
 \Dim=5&A&A&S&S\\
 \hline
 \Dim=6,7&S&A&A&S\\
 \hline
 \Dim=8,9&S&S&A&A\\
 \hline
 \Dim=10,11&A&S&S&A\\
 \hline
 \end{array}
 	\label{dt1} 
 \end{align}
 
(ii) The $\hat\gam\gam_{a_1\ldots a_p}\IC$ are symmetric ($S$) or antisymmetric ($A$) in their spinor indices:
 \begin{align}
 \begin{array}{|l|c|c|c|c|}
 \hline
 &p\Mod 4=0&p\Mod 4=1&p\Mod 4=2&p\Mod 4=3\\
 \hline
 \Dim=4&A&A&S&S\\
 \hline
 \Dim=6&A&A&S&S\\
 \hline
 \Dim=8&S&A&A&S\\
 \hline
 \Dim=10&S&S&A&A\\
 \hline
 \end{array}
 	\label{dt2} 
 \end{align}

(iii) In even dimensions, Weyl spinor bilinears $\psi\gam_{a_1\ldots a_p}\IC\chi^\top$ and $\psi\hat\gam\gam_{a_1\ldots a_p}\IC\chi^\top$ couple either Weyl spinors $\psi,\chi$ of the same chirality ($=$) or Weyl spinors of opposite chiralities ($\neq$):
 \begin{align}
 \begin{array}{|l|c|c|}
 \hline
 &p\Mod 2=0&p\Mod 2=1\\
 \hline
 \Dim=4,8&=&\neq\\
 \hline
 \Dim=6,10&\neq&=\\
 \hline
 \end{array}
 	\label{dt3} 
 \end{align}
 
(iv) For $\Gamma$-matrices and $\CC$ as in \eqref{c12}--\eqref{c13c} one obtains in $\Dim=6,\dots,11$
 \begin{align}
 &\mathrm{for}\ a_1,\ldots,a_p\in\{1,\ldots,\Dim-1\}:\notag
 \\
 &\Dim\in\{6,10\}: 
 \gam_{a_1\ldots a_p}\IC=
 \left\{\begin{array}{rl}
 -\Ii\,\sigma_2\otimes(\gam_{a_1\ldots a_p}\IC)_{(\Dim-1)}& \mathrm{if}\ p\Mod 2=0\\
 \sigma_3\otimes(\gam_{a_1\ldots a_p}\IC)_{(\Dim-1)}& \mathrm{if}\ p\Mod 2=1
 \end{array}\right.
 	\label{dt4}
 	\\
 &\phantom{\Dim\in\{6,10\}:\ } 
 \gam_{\Dim\, a_2\ldots a_p}\IC=
 \Ii\,k_\Dim\left\{\begin{array}{rl}
 \sigma_1\otimes(\gam_{a_2\ldots a_p}\IC)_{(\Dim-1)}& \mathrm{if}\ p\Mod 2=0\\
 -\sigma_0\otimes(\gam_{a_2\ldots a_p}\IC)_{(\Dim-1)}& \mathrm{if}\ p\Mod 2=1
 \end{array}\right.
 	\label{dt5}
 	\\
 	&\Dim=8: 
 \gam_{a_1\ldots a_p}\IC=
 \left\{\begin{array}{rl}
 \sigma_3\otimes(\gam_{a_1\ldots a_p}\IC)_{(7)}& \mathrm{if}\ p\Mod 2=0\\
 -\Ii\,\sigma_2\otimes(\gam_{a_1\ldots a_p}\IC)_{(7)}& \mathrm{if}\ p\Mod 2=1
 \end{array}\right.
 	\label{dt6}
 	\\
 &\phantom{\Dim=8:\ } 
 \gam_{8\, a_2\ldots a_p}\IC=
 \Ii\,k_8\left\{\begin{array}{rl}
 -\sigma_0\otimes(\gam_{a_2\ldots a_p}\IC)_{(7)}& \mathrm{if}\ p\Mod 2=0\\
 \sigma_1\otimes(\gam_{a_2\ldots a_p}\IC)_{(7)}& \mathrm{if}\ p\Mod 2=1
 \end{array}\right.
 	\label{dt7}
 	\\
 	&\Dim\in\{7,9,11\}: 
 \gam_{a_1\ldots a_p}\IC=(\gam_{a_1\ldots a_p}\IC)_{(D-1)}
 	\label{dt8}
 	\\
 &\phantom{\Dim\in\{7,9,11\}:\ } 
 \gam_{\Dim\, a_2\ldots a_p}\IC=
 k_\Dim\,(\hat\gam\gam_{a_2\ldots a_p}\IC)_{(\Dim-1)}
 	\label{dt9}
 \end{align}
where $(\gam_{a_1\ldots a_p}\IC)_{(\Dim-1)}$ and $(\hat\gam\gam_{a_1\ldots a_p}\IC)_{(\Dim-1)}$ denote the matrices $\gam_{a_1\ldots a_p}\IC$ and $\hat\gam\gam_{a_1\ldots a_p}\IC$ in $\Dim-1$ dimensions, respectively.

\section{\texorpdfstring{Primitive elements in $\Dim=4$}{Primitive elements in D=4}}\label{4D}

In $\Dim=4$ one has $\vTHE{1}{\pm}\cdot\vTHE{1}{\pm}=-2\cTHE{11}\mp2$ and $\vTHE{1}{\pm}\cdot\xi^{\pm}_1=\half \THE{11}1$. Lemma 2.9 of ref. \cite{Brandt:2010tz} thus gives:

\begin{lemma}[Primitive elements for $\Ns=1$]\label{lem4D1}\quad\\
For $\Ns=1$ the general solution of the cocycle condition in $\Hg(\fb)$ is:
\begin{align}
  \fb\om=0\ \LRA\ 
  \om\sim&\,
  \cTHE{11}-2 p_-(\xi^-_1)+\cTHE{11}+2 q_+(\xi^+_1)+b\,\THE{11}1\notag\\  
  &+\vTHE{1}{+\ua}\,p_\ua(\xi^-_1)+\vTHE{1}{-\ua}\,q_\ua(\xi^+_1)
  +p_0(\xi^-_1)+q_0(\xi^+_1)
  \label{4D1}
\end{align}
with arbitrary polynomials $p_0(\xi^-_1)$, $p_\ua(\xi^-_1)$, $p_-(\xi^-_1)$ in the components of $\xi^-_1$,
arbitrary polynomials $q_0(\xi^+_1)$, $q_\ua(\xi^+_1)$, $q_+(\xi^+_1)$ in the components of $\xi^+_1$,
and an arbitrary complex number $b\in\mathbb{C}$.
\end{lemma}

Using in addition that in $\Dim=4$ one has $\vTHE{i}{+}\cdot\xi^{+}_j=\cTHE{ij}-1=\cTHE{ji}+1$, lemma 2.10 of ref. \cite{Brandt:2010tz} gives:

\begin{lemma}[Primitive elements for $\Ns=2$]\label{lem4D2}\quad\\
For $\Ns=2$ the general solution of the cocycle condition in $\Hg(\fb)$ is:
\begin{align}
  \fb\om=0\ \LRA\ 
  \om\sim&\,
  \cTHE{12}-1 p_1(\xi^-_1,\xi_2) +\cTHE{12}+1 q_1(\xi^+_1,\xi_2)
  \notag\\
  &+(\THE{11}1-\THE{22}1)\,b(\xi_2)
  +p_0(\xi^-_1,\xi_2)+q_0(\xi^+_1,\xi_2)
  \label{4D2}
\end{align}
with arbitrary polynomials
$p_0(\xi_1^-,\xi_2)$, $p_1(\xi_1^-,\xi_2)$ in the components of $\xi_1^-$ and $\xi_2$, arbitrary polynomials $q_0(\xi_1^+,\xi_2)$, $q_1(\xi_1^+,\xi_2)$ in the components of $\xi_1^+$ and $\xi_2$, and an arbitrary polynomial $b(\xi_2)$ in the components of $\xi_2$.
\end{lemma}

According to lemma 2.11 of ref. \cite{Brandt:2010tz} the cohomology groups $\Hg^p(\fb)$ vanish for all $p>0$ in the cases $\Ns>2$. This gives:

\begin{lemma}[Primitive elements for $\Ns>2$]\label{lem4D3}\quad\\
For $\Ns>2$ the general solution of the cocycle condition in $\Hg(\fb)$ is:
\begin{align}
  \fb\om=0\ \LRA\ 
  \om\sim
  p_0(\xi)
  \label{4D3}
\end{align}
with an arbitrary polynomial $p_0(\xi)$ in the components of $\xi_1,\dots,\xi_\Ns$.
\end{lemma}

%Addition:
{\em Comment:} \\
Lemma \ref{lem4D1} only applies to signatures $(1,3), (2,2), (3,1)$ because
in the cases of signatures $(0,4), (4,0)$ one has $\Ns\in\{2,4,\dots\}$, see \eqref{i4-3}.

\section{\texorpdfstring{Primitive elements in $\Dim=5$}{Primitive elements in D=5}}\label{5D}

%Addition:
The results in $\Dim=5$ are derived by means of the results in $\Dim=4$ as in ref. \cite{Brandt:2010tz} by relating the supersymmetry 
ghosts $\xi_i$ in $\Dim=5$ to supersymmetry 
ghosts $\xi_{i(4)}$ in $\Dim=4$ according to 
\begin{align}
k\in\{1,\dots,\Ns/2\}:\ \xi_{2k-1}\equiv \xi^+_{2k-1(4)}+\xi^-_{2k(4)}\,,\
\xi_{2k}\equiv \xi^+_{2k(4)}-\xi^-_{2k-1(4)}\, ,
  \label{5D0}
\end{align}
and by using \eqref{c12} and \eqref{c12b} which give
\begin{align}
a<5:\ \fb c^a&=\sum_{k=1}^{\Ns/2}\Ii\,\xi_{2k-1}\,\gam^a\IC\,\xi_{2k}^\top
=\Ii\sum_{i=1}^{\Ns}\xi_{i(4)}^+\,(\gam^a\IC)_{(4)}\,\xi_{i(4)}^{-\top}\notag\\
&=\ihalf\sum_{i=1}^{\Ns}\xi_{i(4)}\,(\gam^a\IC)_{(4)}\,\xi_{i(4)}^{\top}
                                =(\fb c^a)_{(4)}\, ,
  \label{5D0a}\\
\fb c^5&=\sum_{k=1}^{\Ns/2}\Ii\,\xi_{2k-1}\,\gam^5\IC\,\xi_{2k}^\top\notag\\
       &=(k_5)^{-1}\sum_{k=1}^{\Ns/2}\Ii\,(\xi^+_{2k-1(4)}\,\IC_{(4)}\,\xi_{2k(4)}^{+\top}
                                +\xi^-_{2k-1(4)}\,\IC_{(4)}\,\xi_{2k(4)}^{-\top})
  \label{5D0b}
\end{align}
where $(\fb c^a)_{(4)}$ denotes $\fb c^a$ in $\Dim=4$.

Lemma 3.1 of ref. \cite{Brandt:2010tz} straightforwardly gives for $\Ns=2$:

\begin{lemma}[Primitive elements for $\Ns=2$]\label{lem5D1}\quad\\
For $\Ns=2$ the general solution of the cocycle condition in $\Hg(\fb)$ is:
\begin{align}
  \fb\om=0\ \LRA\ \om\sim 
	\THE{ij}2\, p_2^{ij}(\xi)+\THE{ij,\, a}2\, p_1^{ij\, a}(\xi)+p_0(\xi)
	\label{5D1}
\end{align}
with arbitrary polynomials $p_0(\xi)$, $p_1^{ij\, a}(\xi)$, $p_2^{ij}(\xi)$ in the components of $\xi_1,\xi_2$.
\end{lemma}

The following result extends and completes lemma 3.2 of ref. \cite{Brandt:2010tz}. It states that
for $\Ns>2$ the cohomology groups $\Hg^p(\fb)$ vanish for all $p>0$:

\begin{lemma}[Primitive elements for $\Ns>2$]\label{lem5D2}\quad\\
For $\Ns>2$ the general solution of the cocycle condition in $\Hg(\fb)$ is:
\begin{align}
  \fb\om=0\ \LRA\ 
  \om\sim
  p_0(\xi)
  \label{5D2}
\end{align}
with an arbitrary polynomial $p_0(\xi)$ in the components of $\xi_1,\dots,\xi_\Ns$.
\end{lemma}

\PF{\ref{lem5D2}}
Because of \eqref{5D0a} $\hHg^p(\fb)$ can be obtained from $\Hg^p(\fb)$ in $\Dim=4$ (for equal values of $\Ns$).
For $\Ns>2$ lemma \ref{lem4D3} implies that $\hHg^p(\fb)$ vanishes for $p\geq 1$. Owing to \eqref{c18} this implies that $\Hg^p(\fb)$ vanishes for $p> 1$,
\begin{align}
 p>1:\quad \fb\om^p=0\ \LRA\ \om^p\sim 0.
 \label{5D6} 
\end{align}
The case $p=1$ is treated by using the results for $\Ns=2$. To this end  
$\fb$ is split into a piece $\fb^{(1)}$ which only involves the supersymmetry ghosts $\xi_1,\xi_2$ and a piece $\fb^{(2)}$ which involves $\xi_3,\dots,\xi_\Ns$:
\begin{align}
\fb=\fb^{(1)}+\fb^{(2)},\quad
\fb^{(1)}=\Ii \,\xi_1\,\gam^a \IC\,\xi^\top_2\,\frac{\6}{\6 c^a}\, .
\label{5D3}
\end{align}
The cocycles $\om^1$ are decomposed into parts $\om^1_m$ with definite degree $m$ in $\xi_1,\xi_2$:
\begin{align}
  \om^1=\sum_{m=0}^{\ol{m}}\om^1_m\,,\quad
  (N_{\xi_1}+N_{\xi_2})\,\om^1_m=m\,\om^1_m
  \label{5D4}
\end{align}
where $N_{\xi_i}$ denotes the counting operator for the components of $\xi_i$. As $\fb^{(1)}$ is the only piece of $\fb$ which increases the degree in $\xi_1,\xi_2$ the part $\om^1_{\ol{m}}$ of $\om^1$ fulfills
\begin{align}
  \fb^{(1)}\om^1_{\ol{m}}=0,\quad \fb^{(2)}\om^1_{\ol{m}}+\fb^{(1)}\om^1_{\ol{m}-2}=0.
  \label{5D5}
\end{align}
The first equation \eqref{5D5} is solved by means of lemma \ref{lem5D1} (with $\xi_3,\dots,\xi_\Ns$ treated as extra variables) since $\fb^{(1)}$ acts as $\fb$ for $\Ns=2$. As we are treating the case $p=1$ we conclude from lemma \ref{lem5D1} that, up to an $\fb^{(1)}$-exact piece, $\om^1_{\ol{m}}$ equals a linear combination of the $\THE{11,\, a}2$, $\THE{12,\, a}2=\THE{21,\, a}2$, $\THE{22,\, a}2$ with coefficients that are polynomials in the supersymmetry ghosts. 
The second equation \eqref{5D5} imposes that $\fb^{(2)}\om^1_{\ol{m}}$ is $\fb^{(1)}$-exact. From this condition one derives that $\om^1_{\ol{m}}$ is $\fb^{(1)}$-exact by itself by exploiting $\fb^{(2)}\THE{11,\, a}2$, $\fb^{(2)}\THE{12,\, a}2$, $\fb^{(2)}\THE{22,\, a}2$ \cite{note1}.
This implies that $\om^1_{\ol{m}}$ can be removed from $\om^1$ by subtracting a coboundary in $\Hg(\fb)$ from $\om^1$. In the same way all other parts $\om^1_m$ can be successively removed from $\om^1$ implying that $\om^1$ is a coboundary in $\Hg(\fb)$,
\begin{align}
 \fb\om^1=0\ \LRA\ \om^1\sim 0.
 \label{5D7} 
\end{align}
The lemma is obtained from \eqref{5D6}, \eqref{5D7} and $\om^0=p_0(\xi)$ (which trivially fulfills $\fb\om^0=0$). 
\QED

\section{\texorpdfstring{Primitive elements in $\Dim=6$}{Primitive elements in D=6}}\label{6D}

\subsection{Signatures (1,5), (3,3), (5,1)}\label{6D.1}

In the cases of signatures $(1,5),(3,3),(5,1)$ the supersymmetry ghosts $\xi_i$ are Majorana Weyl spinors (for signature $(3,3)$) or symplectic Majorana Weyl spinors (for signatures $(1,5),(5,1)$). $\Ns_+$ denotes the number of supersymmetry ghosts with positive chirality, $\Ns_-$ denotes the number of supersymmetry ghosts with negative chirality. Both $\Ns_+$ and $\Ns_-$ are even integers, $\Ns_+,\Ns_-\in\{0,2,\dots\}$. $\Ns$ is the sum  $\Ns=\Ns_++\Ns_-\in\{2,4\dots\}$. The case $\Ns=2$ thus includes $(\Ns_+,\Ns_-)=(2,0)$ and $(\Ns_+,\Ns_-)=(0,2)$, the case $\Ns=4$ includes $(\Ns_+,\Ns_-)=(4,0)$, $(\Ns_+,\Ns_-)=(2,2)$ and $(\Ns_+,\Ns_-)=(0,4)$ etc. We use $\xi_i=\xi_i^+$ for $i\leq\Ns_+$ and $\xi_i=\xi_i^-$ for $i>\Ns_+$, i.e. the supersymmetry ghosts $\xi_1,\dots,\xi_{\Ns_+}$ have positive chirality and the supersymmetry ghosts $\xi_{\Ns_++1},\dots,\xi_{\Ns_++\Ns_-}$ have negative chirality.

In a spinor representation fulfilling \eqref{c13} and \eqref{c13b} a Weyl spinor $\psi^+=\psi^+\hat\gam$ with positive chirality takes the form $\psi^+=(\chi,0)$ and a Weyl spinor $\psi^-=-\psi^-\hat\gam$ with negative chirality takes the form $\psi^-=(0,\chi)$ where $\chi$ and $0$ have four components, respectively, like spinors in $\Dim=5$. In order to derive $\Hg(\fb)$ in $\Dim=6$ by means of $\Hg(\fb)$ in $\Dim=5$, we relate the supersymmetry ghosts $\xi_i$ in $\Dim=6$ to supersymmetry ghosts $\xi_{i(5)}$ in $\Dim=5$ as follows:
\begin{align}
  i\leq \Ns_+:&\ \xi_i\equiv (\xi_{i(5)},0),\quad 
  i>\Ns_+:\ \xi_i\equiv (0,\Ii\,\xi_{i(5)}).\label{6D6}
\end{align}
\eqref{dt4} and \eqref{dt5} for $p=1$ and \eqref{6D6} give:
\begin{align}
a<6:\ &\fb c^a=\sum_{k=1}^{\Ns/2}\Ii\,\xi_{2k-1}\,\gam^a\IC\,\xi_{2k}^\top
                                =\sum_{k=1}^{\Ns/2}\Ii\,\xi_{2k-1(5)}\,(\gam^a\IC)_{(5)}\,\xi_{2k(5)}^\top
                                =(\fb c^a)_{(5)}\, ,
  \label{6D7}
  \\
  &\fb c^6=(k_6)^{-1}\Big(\sum_{k=1}^{\Ns_+/2}\xi_{2k-1(5)}\,\IC_{(5)}\,\xi_{2k(5)}^\top
  -\sum_{k=\Ns_+/2+1}^{\Ns/2}\xi_{2k-1(5)}\,\IC_{(5)}\,\xi_{2k(5)}^\top\Big) .
  \label{6D8}
\end{align}
where $\Ns=\Ns_++\Ns_-$, and $(\fb c^a)_{(5)}$ denotes $\fb c^a$ in $\Dim=5$. Hence, using a spinor representation fulfilling \eqref{c13} and \eqref{c13b} and the identifications \eqref{6D6}, the action of $\fb$ in $\hOmg$ in $\Dim=6$ is identical to the action of $\fb$ in $\Omg$ in $\Dim=5$. This is used to derive $\hHg(\fb)$ in $\Dim=6$ by means of the results for $\Hg(\fb)$ in $\Dim=5$.

\begin{lemma}[Primitive elements for $\Ns_++\Ns_-=2$]\label{lem6D1}\quad \\
In the cases $(\Ns_+,\Ns_-)=(2,0)$ and $(\Ns_+,\Ns_-)=(0,2)$ the general solution of the cocycle condition in $\Hg(\fb)$ is:
\begin{align}
 \fb\om=0\ \LRA\ \om\sim
 \THE{ij}3\, p_3^{ij}(\xi)+\THE{ij,\, a}3\, p_2^{ij\, a}(\xi)
 +\vTHE{i}{\ua}\, p^i_\ua(\xi)
 +p_0(\xi)
 \label{6D1} 
\end{align}
with arbitrary polynomials $p_0(\xi)$, $p^i_\ua(\xi)$, $p_2^{ij\, a}(\xi)$, $p_3^{ij}(\xi)$ in the components of $\xi_1,\xi_2$.
\end{lemma}

{\bf Proof of lemma \ref{lem6D1}:}
As outlined above, $\hHg^p(\fb)$ is obtained from $\Hg^p(\fb)$ in $\Dim=5$ using a spinor representation fulfilling \eqref{c13} and \eqref{c13b} and the identifications \eqref{6D6}. Lemma \ref{lem5D1} implies that $\hHg^p(\fb)$ vanishes for $p\geq 3$. Owing to \eqref{c18} this implies that $\Hg^p(\fb)$ vanishes for $p> 3$,
\begin{align}
 p>3:\quad \fb\om^p=0\ \LRA\ \om^p\sim 0.
 \label{6D9} 
\end{align}
In the case $p=3$ \eqref{c8}, \eqref{c10} and lemma \ref{lem5D1} imply that we may assume
\begin{align}
 p=3:\quad \hat\om^2=\THE{ij(5)}2\, \hat p_3^{ij}(\xi_{(5)}),\quad \hat\om^3_{\mathrm{hom}}=0
 \label{6D10} 
\end{align}
where $\hat p_3^{ij}(\xi_{(5)})$ are polynomials in $\xi_{1(5)},\xi_{2(5)}$, and $\THE{ij(5)}2$ denotes $\THE{ij}2$ in $\Dim=5$. 
\eqref{def2},
\eqref{dt5} for $p=3$ and \eqref{6D6} give for $(\Ns_+,\Ns_-)=(2,0)$ and $(\Ns_+,\Ns_-)=(0,2)$:
\begin{align}
 \THE{ij}3\propto c^6\,\THE{ij(5)}2+\ldots
 \label{6D11} 
\end{align}
where ellipses indicate terms without $c^6$. $\THE{ij}3$ is a cocycle in $\Hg^3(\fb)$ \cite{note2a}.
Hence, $c^6\hat\om^2$ can be completed to the cocycle $\THE{ij}3 p_3^{ij}(\xi)$ (with $p_3^{ij}(\xi)\propto \hat p_3^{ij}(\xi_{(5)})$).
 This gives
\begin{align}
 \fb\om^3=0\ \LRA\ \om^3\sim \THE{ij}3\, p_3^{ij}(\xi).
 \label{6D12} 
\end{align}
In the case $p=2$ \eqref{c8}, \eqref{c10} and lemma \ref{lem5D1} imply that we may assume
\begin{align}
 p=2:\quad \hat\om^1=\THE{ij,\, a(5)}2\, \hat p_2^{ij\, a}(\xi_{(5)}),\ \hat\om^2_{\mathrm{hom}}=\THE{ij(5)}2\, \hat p_2^{ij\, 6}(\xi_{(5)})
 \label{6D13} 
\end{align}
where the sum over $a$ runs from $a=1$ to $a=5$ (as there is no $\THE{ij,\, 6}2$ in $\Dim=5$).
Using \eqref{6D11} one obtains for $(\Ns_+,\Ns_-)=(2,0)$ and $(\Ns_+,\Ns_-)=(0,2)$:
\begin{align}
 a<6:\ \THE{ij,\, a}3= \frac{\6 \THE{ij}3}{\6 c^a}\propto c^6\, \THE{ij,\, a(5)}2+\dots,\quad
\THE{ij,\, 6}3=\frac{\6 \THE{ij}3}{\6 c^6}\propto\THE{ij(5)}2
 \label{6D14} 
\end{align}
where again ellipses indicate terms without $c^6$. According to the first equation \eqref{6D14} $c^6\hat\om^1$ can be completed to
$\sum_{a=1}^5\THE{ij,\, a}3\, p_2^{ij\, a}(\xi)$ which is a cocycle in $\Hg^2(\fb)$  because $\fb\THE{ij}3=0$ implies $\fb\THE{ij,\,a}3=0$, see \eqref{c19}. Using in addition the second equation \eqref{6D14}
we conclude
\begin{align}
 \fb\om^2=0\ \LRA\ \om^2\sim \THE{ij,\, a}3\, p_2^{ij\, a}(\xi)
 \label{6D15}
\end{align}
where here the sum over $a$ runs from $a=1$ to $a=6$.
 
In the case $p=1$ we use that every cocycle $\om^1$ must be at least linear in the supersymmetry ghosts since no nonvanishing linear combination of the $c^a$ with constant coefficients is $\fb$-closed. This yields, along with  lemma \ref{lem5D1} for $p=1$:
\begin{align}
 p=1:\quad \hat\om^0=\xi_{i(5)}^\ua\,\hat p^{i}_\ua(\xi_{(5)}),\ \hat\om^1_{\mathrm{hom}}=\THE{ij,\, a(5)}2\, \hat p_1^{ij\, 6\,a}(\xi_{(5)}).
 \label{6D16} 
\end{align}
Every monomial $c^6\xi_{i(5)}^\ua$ corresponds to a cocycle in $\Hg^1(\fb)$ proportional to one of the $\vTHE{i}{\ua}$ \cite{note2b}. Using additionally $\THE{ij,\, a(5)}2\propto\vTHE {(i}{}\,\gam_{6a}\IC\,\xi_{j)}^\top$ \cite{note2} this yields
\begin{align}
 \fb\om^1=0\ \LRA\ \om^1\sim \vTHE{i}{\ua}\, p^i_\ua(\xi).
 \label{6D17}
\end{align}
The lemma is obtained from \eqref{6D9}, \eqref{6D12}, \eqref{6D15}, \eqref{6D17} and $\om^0=p_0(\xi)$. \QED

\begin{lemma}[Primitive elements for $\Ns_++\Ns_-=4$]\label{lem6D2}\quad \\
(i) In the case $(\Ns_+,\Ns_-)=(2,2)$ the general solution of the cocycle condition in $\Hg(\fb)$ is:
\begin{align}
 \fb\om=0\ \LRA\ \om\sim
 (\THE{12}1-\THE{34}1)\, p_1(\xi)+p_0(\xi)
 \label{6D2a} 
\end{align}
with arbitrary polynomials
$p_0(\xi)$, $p_1(\xi)$ in the components of $\xi_1,\dots,\xi_4$.

(ii) In the cases $(\Ns_+,\Ns_-)=(4,0)$ and $(\Ns_+,\Ns_-)=(0,4)$ the general solution of the cocycle condition in $\Hg(\fb)$ is:
\begin{align}
 \fb\om=0\ \LRA\ \om\sim
 &\ \THE{13}1\, p^{13}_1(\xi)
 +\THE{14}1\, p^{14}_1(\xi)+\THE{23}1\, p^{23}_1(\xi)\notag\\
 &
 +\THE{24}1\, p^{24}_1(\xi)+(\THE{12}1-\THE{34}1)\, p_1(\xi)+p_0(\xi)
 \label{6D2} 
\end{align}
with arbitrary polynomials
$p_0(\xi)$, $p_1(\xi)$, $p^{24}_1(\xi)$, $p^{23}_1(\xi)$, $p^{14}_1(\xi)$, $p^{13}_1(\xi)$ in the components of $\xi_1,\dots,\xi_4$.
\end{lemma}

\PF{\ref{lem6D2}}
Lemma \ref{lem5D2} implies that $\hHg^p(\fb)$ vanishes for  all $p\geq 1$ in the cases $\Ns_++\Ns_-= 4$. Owing to \eqref{c18} this implies that $\Hg^p(\fb)$ vanishes for all $p> 1$,
\begin{align}
 p>1:\quad \fb\om^p=0\ \LRA\ \om^p\sim 0.
 \label{6D18} 
\end{align} 
The case $p=1$ is analyzed analogously to the case $p=1$ in the proof sketch for lemma \ref{lem5D2} by using the decompositions \eqref{5D3} and \eqref{5D4} of $\fb$ and $\om^1$, leading to equations \eqref{5D5}. The first equation \eqref{5D5} is solved by means of lemma \ref{lem6D1} which for $\ol m>0$ gives 
$\om^1_{\ol{m}}=\sum_{i=1}^2\vTHE i{\ua}\,p^i_\ua(\xi)$ for polynomials $p^1_\ua(\xi)$, $p^2_\ua(\xi)$, up to an $\fb^{(1)}$-exact piece which may be neglected. The
$\fb^{(2)}\vTHE 1\ua$ and $\fb^{(2)}\vTHE 2\ua$ are linear in $\xi_1, \xi_2$ and linearly independent. Therefore, no
nonvanishing linear combination (with constant coefficients) of the $\fb^{(2)}\vTHE 1\ua$ and $\fb^{(2)}\vTHE 2\ua$ is $\fb^{(1)}$-exact. The second equation \eqref{5D5} thus implies that the polynomials $p^1_\ua(\xi)$ and $p^2_\ua(\xi)$ are at least linear in the supersymmetry ghosts.

(i) In the case $(\Ns_+,\Ns_-)=(2,2)$ the second equation \eqref{5D5} further implies that, up to an $\fb^{(1)}$-exact piece which may be neglected,
$\sum_{i=1}^2\vTHE i{\ua}\,p^i_\ua(\xi)$ equals $\THE{12}1=\vTHE 1{}\cdot \xi_2=-\vTHE 2{}\cdot \xi_1$ times a polynomial $p_{1\,\ol{m}-2}(\xi)$. $\THE{12}1$ is completed to the cocycle $\THE{12}1-\THE{34}1$ in $\Hg^1(\fb)$ and $\om^{1\,\prime}=\om^1-(\THE{12}1-\THE{34}1)\,p_{1\,\ol{m}-2}(\xi)$ can now be treated as $\om^1$ before. Repeating the arguments one obtains:
\begin{align}
 (\Ns_+,\Ns_-)=(2,2):\ \fb\om^1=0\ \LRA\ 
 \om^1\sim 
 (\THE{12}1-\THE{34}1)\, p_1(\xi).
 \label{6D19a} 
\end{align} 

(ii) In the cases $(\Ns_+,\Ns_-)=(4,0)$ and $(\Ns_+,\Ns_-)=(0,4)$
the second equation \eqref{5D5} implies that, up to an $\fb^{(1)}$-exact piece which may be neglected, $\sum_{i=1}^2\vTHE i{\ua}\,p^i_\ua(\xi)$ equals a linear combination of $\THE{12}1$, $\THE{13}1$, $\THE{14}1$, $\THE{23}1$, $\THE{24}1$ with coefficients that are polynomials in the components of the supersymmetry ghosts. $\THE{13}1$, $\THE{14}1$, $\THE{23}1$, $\THE{24}1$ are cocycles in $\Hg^1(\fb)$ by themselves, $\THE{12}1$ is completed to the cocycle $\THE{12}1-\THE{34}1$ in $\Hg^1(\fb)$. Proceeding now analogously to case (i) one obtains:
\begin{align}
&(\Ns_+,\Ns_-)\in\{(4,0),(0,4)\}:\ 
 \fb\om^1=0\ \LRA\notag\\ 
 &\om^1\sim 
 \THE{13}1\, p^{13}_1(\xi)
 +\THE{14}1\, p^{14}_1(\xi)+\THE{23}1\, p^{23}_1(\xi)
 +\THE{24}1\, p^{24}_1(\xi)+(\THE{12}1-\THE{34}1)\, p_1(\xi).
 \label{6D19} 
\end{align} 
The lemma is obtained from \eqref{6D18}, \eqref{6D19a}, \eqref{6D19} and $\om^0=p_0(\xi)$. \QED

\begin{lemma}[Primitive elements for $\Ns_++\Ns_->4$]\label{lem6D3}\quad \\
In the cases $\Ns_++\Ns_->4$ the general solution of the cocycle condition in $\Hg(\fb)$ is:
\begin{align}
 \fb\om=0\ \LRA\ \om\sim
 p_0(\xi)
 \label{6D3} 
\end{align}
with an arbitrary polynomial $p_0(\xi)$ in the components of $\xi_1,\dots,\xi_\Ns$.
\end{lemma}

\PF{\ref{lem6D3}}
As in the proof of lemma \ref{lem6D2} one concludes that $\Hg^p(\fb)$ vanishes for all $p> 1$.
The case $p=1$ is analyzed by using decompositions of $\fb$ and $\om^1$ similar to \eqref{5D3} and \eqref{5D4}, but now with a piece $\fb^{(1)}$ of $\fb$ which involves $\xi_1,\dots,\xi_4$ (in place of only $\xi_1,\xi_2$) and parts $\om_m^1$ of $\om^1$ which have degree $m$ in $\xi_1,\dots,\xi_4$. This leads again to equations \eqref{5D5}. The first equation \eqref{5D5} is solved by means of lemma \ref{lem6D2} which gives 
$\om^1_{\ol{m}}=(\THE{12}1-\THE{34}1)\, p_{1\,\ol{m}-2}(\xi)+\fb^{(1)}(\dots)$ (for $\Ns_+=2$) or $\om^1_{\ol{m}}=\THE{13}1\, p^{13}_{1\,\ol{m}-1}(\xi)+\dots+(\THE{12}1-\THE{34}1)\, p_{1\,\ol{m}-2}(\xi)+\fb^{(1)}(\dots)$ (for $\Ns_+\neq 2$).
The second equation \eqref{5D5} then implies in either case that $\om^1_{\ol{m}}$ is $\fb^{(1)}$-exact and can thus be removed from $\om^1$ by subtracting a coboundary in $\Hg(\fb)$. Repeating the arguments one concludes that all other parts $\om_m^1$ can be removed in the same way which gives $\om^1\sim 0$. $p=0$ gives $\om^0=p_0(\xi)$. \QED

{\em Comments:} \\
1. The difference between the results for $(\Ns_+,\Ns_-)=(2,2)$ and for $(\Ns_+,\Ns_-)\in\{(4,0),$ $(0,4)\}$ parallels the situation in $\Dim=2$ \cite{Brandt:2010fa} and $\Dim=10$.

%Addition:
2. We note that in lemma \ref{lem6D1} one has $\THE{ij}3=\cTHE{ij}+3$ and $\vTHE{i}{}=\vTHE{i}-$ in the case $(\Ns_+,\Ns_-)=(2,0)$, and $\THE{ij}3=\cTHE{ij}-3$ and $\vTHE{i}{}=\vTHE{i}+$ in the case $(\Ns_+,\Ns_-)=(0,2)$.
Analogously in lemma \ref{lem6D2} one has $\THE{ij}1=\cTHE{ij}+1$ in the case $(\Ns_+,\Ns_-)=(4,0)$, and $\THE{ij}1=\cTHE{ij}-1$ in the case $(\Ns_+,\Ns_-)=(0,4)$.

\subsection{Signatures (0,6), (2,4), (4,2), (6,0)}\label{6D.2}

In the cases of signatures $(0,6),(2,4),(4,2),(6,0)$ the supersymmetry ghosts $\xi_i$ are Majorana spinors consisting of two Weyl spinors with opposite chiralities, respectively. $\Ns\in\{2,4,\dots\}$ denotes the number of these Majorana supersymmetry ghosts. Hence, there are both $\Ns$ Weyl supersymmetry ghosts with positive chirality and $\Ns$ Weyl supersymmetry ghosts with negative chirality. 

The case $\Ns=2$ corresponds thus to the case $(\Ns_+,\Ns_-)=(2,2)$ in lemma \ref{lem6D2}. Using $\xi_i\,\hat\gam=\xi^+_i-\xi^-_i$ and identifying $\xi^+_1,\xi^+_2,\xi^-_1,\xi^-_2$ with $\xi_1,\xi_2,\xi_3,\xi_4$ in lemma \ref{lem6D2}, respectively, lemma \ref{lem6D2} gives directly:

\begin{lemma}[Primitive elements for $\Ns=2$]\label{lem6D4}\quad \\
In the case $\Ns=2$ the general solution of the cocycle condition in $\Hg(\fb)$ is:
\begin{align}
 \fb\om=0\ \LRA\ \om\sim
 \hTHE{12}1\, p_1(\xi)+p_0(\xi)
 \label{6D4} 
\end{align}
with arbitrary polynomials $p_0(\xi)$, $p_1(\xi)$ in the components of $\xi_1,\xi_2$.
\end{lemma}

The cases $\Ns>2$ correspond to cases $\Ns_+=\Ns_->2$ in lemma \ref{lem6D3} which implies: 

\begin{lemma}[Primitive elements for $\Ns>2$]\label{lem6D5}\quad \\
In the cases $\Ns>2$ the general solution of the cocycle condition in $\Hg(\fb)$ is:
\begin{align}
 \fb\om=0\ \LRA\ \om\sim
 p_0(\xi)
 \label{6D5} 
\end{align}
with an arbitrary polynomial $p_0(\xi)$ in the components of $\xi_1,\dots,\xi_\Ns$.
\end{lemma}

\section{\texorpdfstring{Primitive elements in $\Dim=7$}{Primitive elements in D=7}}\label{7D}

The results in $\Dim=7$ are derived by means of the results in $\Dim=6$
presented in section \ref{6D.2}. To this end we use \eqref{c12} and \eqref{c12a} which give
\begin{align}
a<7:\ &\fb c^a=\sum_{k=1}^{\Ns/2}\Ii\,\xi_{2k-1}\,\gam^a\IC\,\xi_{2k}^\top
                                =\sum_{k=1}^{\Ns/2}\Ii\,\xi_{2k-1(6)}\,(\gam^a\IC)_{(6)}\,\xi_{2k(6)}^\top
                                =(\fb c^a)_{(6)}\, ,
  \label{7D3}\\
&\fb c^7=\sum_{k=1}^{\Ns/2}\Ii\,\xi_{2k-1}\,\gam^7\IC\,\xi_{2k}^\top
                                =(k_7)^{-1}\sum_{k=1}^{\Ns/2}\Ii\,\xi_{2k-1(6)}\,(\hat\gam\IC)_{(6)}\,\xi_{2k(6)}^\top
  \label{7D4}
\end{align}
where $(\fb c^a)_{(6)}$ and $\xi_{i(6)}$ denote $\fb c^a$ and $\xi_i$ in $\Dim=6$, each $\xi_{i(6)}$ being an 8-component spinor consisting of two Weyl spinors, with 
$\xi_i\equiv \xi_{i(6)}$. \eqref{7D3} shows that we can use lemmas \ref{lem6D4} and \ref{lem6D5} to obtain $\hHg(\fb)$ in $\Dim=7$.
 
\begin{lemma}[Primitive elements for $\Ns=2$]\label{lem7D1}\quad \\
In the case $\Ns=2$ the general solution of the cocycle condition in $\Hg(\fb)$ is:
\begin{align}
 \fb\om=0\ \LRA\ \om\sim
 \THE{12}2\, p_2(\xi)+\THE{12,\, a}2\, p^a_1(\xi)
 +p_0(\xi)
 \label{7D1} 
\end{align}
with arbitrary polynomials $p_0(\xi)$, $p^a_1(\xi)$, $p_2(\xi)$ in the components of $\xi_1,\xi_2$.
\end{lemma}

\PF{\ref{lem7D1}}
Lemma \ref{lem6D4} implies that $\hHg^p(\fb)$ vanishes for $p\geq 2$. Using \eqref{c18} we conclude that $\Hg^p(\fb)$ vanishes for $p> 2$,
\begin{align}
 p>2:\quad \fb\om^p=0\ \LRA\ \om^p\sim 0.
 \label{7D6} 
\end{align}
For $p=2$ and $p=1$ the following implications of \eqref{c12} and \eqref{c12a} are used:
\begin{align}
 &\THE{12}2\propto c^7\,\hTHE{12(6)}1+\ldots\ ,\label{7D8}
 \\
  a<7:\ &\THE{12,\, a}2=\frac{\6 \THE{12}2}{\6 c^a}\propto c^7\,\frac{\6 \hTHE{12(6)}1}{\6 c^a}+\ldots=
 c^7\,\xi_{1(6)}\,(\hat\gam\gam_a\IC)_{(6)}\, \xi_{2(6)}^\top+\ldots\ ,\label{7D8a}
 \\
 &\THE{12,\, 7}2=\frac{\6 \THE{12}2}{\6 c^7}\propto\hTHE{12(6)}1
 \label{7D8b} 
\end{align}
where ellipses indicate terms without $c^7$. 

Using lemma \ref{lem6D4} and exploiting \eqref{c7} in the case $p=1$ \cite{note3} one obtains that one may assume:
\begin{align}
 p=2:\ &\hat\om^1=\hTHE{12(6)}1\, \hat p_2(\xi_{(6)}),\quad \hat\om^2_{\mathrm{hom}}=0;
 \label{7D7} \\
 p=1:\ &\hat\om^0=\xi_{1(6)}\,(\hat\gam\gam_a\IC)_{(6)}\, \xi_{2(6)}^\top\,\hat p^a_1(\xi_{(6)}),\ \hat\om^1_{\mathrm{hom}}=\hTHE{12(6)}1\, \hat p^7_1(\xi_{(6)}).\label{7D10}
\end{align}
Using \eqref{7D8}--\eqref{7D8b} and $\fb\THE{12}2=0$ which may be verified explicitly one obtains:
\begin{align}
 \fb\om^2=0\ \LRA\ &\om^2\sim \THE{12}2\, p_2(\xi),\label{7D9} 
 \\
 \fb\om^1=0\ \LRA\ &\om^1\sim \THE{12,\, a}2\, p^a_1(\xi).
 \label{7D12}
\end{align} 
The lemma is obtained from \eqref{7D6}, \eqref{7D9}, \eqref{7D12} and $\om^0=p_0(\xi)$. \QED

The cases $\Ns>2$ can be analyzed analogously to $\Ns>2$ in $\Dim=5$, see proof sketch for lemma \ref{lem5D2}, which gives:

\begin{lemma}[Primitive elements for $\Ns>2$]\label{lem7D2}\quad \\
In the cases $\Ns>2$ the general solution of the cocycle condition in $\Hg(\fb)$ is:
\begin{align}
 \fb\om=0\ \LRA\ \om\sim p_0(\xi)
 \label{7D2} 
\end{align}
with an arbitrary polynomial $p_0(\xi)$ in the components of $\xi_1,\dots,\xi_\Ns$.
\end{lemma}

\section{\texorpdfstring{Primitive elements in $\Dim=8$}{Primitive elements in D=8}}\label{8D}

In order to derive $\Hg(\fb)$ in $\Dim=8$ by means of $\Hg(\fb)$ in $\Dim=7$, we 
use in $\Dim=8$ a spinor representation fulfilling \eqref{c13} and \eqref{c13a} and
relate the 16-component supersymmetry ghosts $\xi_i$ in $\Dim=8$ to 8-component supersymmetry ghosts $\xi_{i(7)}$ in $\Dim=7$ as follows:
\begin{align}
  \xi_i\equiv \Ii\,(\xi_{2i-1(7)},\xi_{2i(7)}).\label{8D1}
\end{align}
Using \eqref{dt6} and \eqref{dt7} for $p=1$ these identifications give:
\begin{align}
a<8:\quad \fb c^a&=\half\,\sum_{i=1}^\Ns\xi_i\, [\sigma_2\otimes(\gam^a\IC)_{(7)}]  \,\xi_i^\top\notag\\
&=\sum_{k=1}^{\Ns}\Ii\,\xi_{2k-1(7)}\,(\gam^a\IC)_{(7)}\,\xi_{2k(7)}^\top
                                =(\fb c^a)_{(7)}\, ,
  \label{8D2}\\
  \fb c^8&=(k_8)^{-1}\sum_{k=1}^{\Ns}\xi_{2k-1(7)}\,\IC_{(7)}\,\xi_{2k(7)}^\top
  \label{8D2a}
\end{align}
where $(\fb c^a)_{(7)}$ denotes $\fb c^a$ in $\Dim=7$. Hence, using a spinor representation fulfilling \eqref{c13} and \eqref{c13a} and the identifications \eqref{8D1}, the action of $\fb$ in $\hOmg$ in $\Dim=8$ for $\Ns$ supersymmetry ghosts $\xi_i$ is identical to the action of $\fb$ in $\Omg$ in $\Dim=7$ for $2\Ns$ supersymmetry ghosts $\xi_{i(7)}$. 
This is used to derive $\hHg(\fb)$ in $\Dim=8$ by means of $\Hg(\fb)$ in $\Dim=7$.

\begin{lemma}[Primitive elements for $\Ns=1$]\label{lem8D1}\quad \\
In the case $\Ns=1$ the general solution of the cocycle condition in $\Hg(\fb)$ is:
\begin{align}
 \fb\om=0\ \LRA\ \om\sim
 \hTHE{11}3\, p_3(\xi)+\hTHE{11,\, a}3\, p^a_2(\xi)+\hTHE{11,\, ab}3\, p^{ab}_1(\xi)
 +p_0(\xi)
 \label{8D3} 
\end{align}
with arbitrary polynomials $p_0(\xi)$, $p^{ab}_1(\xi)$, $p^a_2(\xi)$, $p_3(\xi)$ in the components of $\xi_1$.
\end{lemma}

\PF{\ref{lem8D1}}
Lemma \ref{lem7D1} implies that $\hHg^p(\fb)$ vanishes for $p\geq 3$. Using \eqref{c18} we conclude that $\Hg^p(\fb)$ vanishes for $p> 3$,
\begin{align}
 p>3:\quad \fb\om^p=0\ \LRA\ \om^p\sim 0.
 \label{8D4} 
\end{align}
For $3\geq p\geq 1$ the following implications of \eqref{c13} and \eqref{c13a} are used:
\begin{align}
 &\hTHE{11}3\propto c^8\,\THE{12(7)}2+\ldots\ ,\label{8D6a}
 \\
  a<8:\ &\hTHE{11,\, a}3=\frac{\6 \hTHE{11}3}{\6 c^a}\propto c^8\,\THE{12,\, a(7)}2+\ldots\ ,\label{8D6b}
 \\
 &\hTHE{11,\, 8}3=\frac{\6 \hTHE{11}3}{\6 c^8}\propto\THE{12(7)}2\, ,\label{8D6c}
 \\
 a,b<8:\ &\hTHE{11,\, ab}3=\frac{\6^2 \hTHE{11}3}{\6 c^b\6 c^a}\propto c^8\,\xi_{1(7)}\,(\gam_{ab}\IC)_{(7)}\,\xi_{2(7)}^\top+\ldots\ ,\label{8D6d}
 \\
 a<8:\ &\hTHE{11,\, 8\,a}3=\frac{\6^2 \hTHE{11}3}{\6 c^a\6 c^8}\propto \THE{12,\, a(7)}2\label{8D6e}
\end{align}
where ellipses indicate terms without $c^8$.

Using lemma \ref{lem7D1} and exploiting \eqref{c7} in the case $p=1$ \cite{note4} one obtains that one may assume:
\begin{align}
 p=3:\ &\hat\om^2=\THE{12(7)}2\, \hat p_3(\xi_{(7)}),\ \hat\om_\mathrm{hom}^3=0;\label{8D5a}
 \\
 p=2:\ &\hat\om^1=\THE{12,\, a(7)}2\, \hat p^a_2(\xi_{(7)}),\ \hat\om_\mathrm{hom}^2=\THE{12(7)}2\,\hat p^8_2(\xi_{(7)});\label{8D5b}
 \\
 p=1:\ &\hat\om^0=\xi_{1(7)}\,(\gam_{ab}\IC)_{(7)}\,\xi_{2(7)}^\top\, \hat p^{ab}_1(\xi_{(7)}),\ 
        \hat\om_\mathrm{hom}^1=\THE{12,\, a(7)}2\, 
        \hat p^{8\,a}_1(\xi_{(7)}).\label{8D5c}
\end{align}
Using \eqref{8D6a}--\eqref{8D6e} and $\fb\hTHE{11}3=0$ which may be verified explicitly one obtains:
\begin{align}
 \fb\om^3=0\ &\LRA\ \om^3\sim\hTHE{11}3\, p_3(\xi),\label{8D7a}
 \\
 \fb\om^2=0\ &\LRA\ \om^2\sim\hTHE{11,\, a}3\, p^a_2(\xi),\label{8D7b}
 \\
 \fb\om^1=0\ &\LRA\ \om^1\sim\hTHE{11,\, ab}3\, p^{ab}_1(\xi).\label{8D7c}
\end{align}
The lemma is obtained from \eqref{8D4}, \eqref{8D7a}--\eqref{8D7c} and $\om^0=p_0(\xi)$.\QED

\begin{lemma}[Primitive elements for $\Ns>1$]\label{lem8D2}\quad \\
In the cases $\Ns>1$ the general solution of the cocycle condition in $\Hg(\fb)$ is:
\begin{align}
 \fb\om=0\ \LRA\ \om\sim
 p_0(\xi)
 \label{8D8} 
\end{align}
with an arbitrary polynomial $p_0(\xi)$ in the components of $\xi_1,\dots,\xi_\Ns$.
\end{lemma}

\PF{\ref{lem8D2}}
Lemma \ref{lem7D2} implies that $\hHg^p(\fb)$ vanishes for all $p\geq 1$ in the cases $\Ns> 1$. Using \eqref{c18} one infers that $\Hg^p(\fb)$ vanishes for all $p> 1$.
The case $p=1$ is analyzed using decompositions of $\fb$ and $\om^1$ similar to \eqref{5D3} and \eqref{5D4}, but now with a piece $\fb^{(1)}=\ihalf\,\xi_1\, \gam^a\IC \,\xi_1^\top\,\6/\6 c^a$ of $\fb$ and parts $\om_m^1$ of $\om^1$ with degree $m$ in $\xi_1$. Using lemma \ref{lem8D1} one infers from the first equation 
\eqref{5D5} that
$\om^1_{\ol{m}}=\hTHE{11,\, ab}3\, p^{ab}(\xi)+\fb^{(1)}(\dots)$ for polynomials $p^{ab}(\xi)$ in $\xi_1,\dots,\xi_\Ns$.
The second equation \eqref{5D5} then implies that the $p^{ab}(\xi)$ are such that $\om^1_{\ol{m}}$ is $\fb^{(1)}$-exact and can thus be removed from $\om^1$ by subtracting a coboundary in $\Hg(\fb)$. Repeating the arguments one concludes that all other parts $\om_m^1$ can be removed in the same way which gives $\om^1\sim 0$. $p=0$ gives $\om^0=p_0(\xi)$. \QED

{\em Comments:} \\
1. It may be noted that in $\Dim=8$ there are neither primitive elements $\vTHE i{}$ or $\vTHE i\pm$ nor primitive elements $\THE {ij}{\Dim/2}$, $\hTHE {ij}{\Dim/2}$ or $\cTHE {ij}\pm{\Dim/2}$, in contrast to $\Dim=4,6,10$. This result in $\Dim=8$ may be traced back to the more general feature that in $\Dim=8$  there are no cocycles at all in $\Hg^p(\fb)$ for $p>0$ which depend only on components of chiral spinors $\xi_i^+$ of positive chirality or only on components of chiral spinors $\xi_i^-$ with negative chirality, again in contrast to $\Dim=4,6,10$. This feature can be proved as follows for $\Ns=1$ (and analogously for $\Ns>1$) in a spinor representation fulfilling \eqref{c13} and \eqref{c13a}, using \eqref{8D1}. 
The cocycle condition $\fb\om^p=0$ can for $\Ns=1$ be written as 
\begin{align}
\xi_{1(7)}^\ua\,R_\ua{}^a\ \frac{\6\om^p}{\6 c^a}=0
\label{8D9} 
\end{align}
where according to \eqref{8D2} and \eqref{8D2a} $R_\ua{}^a$ are the entries of an $8\times 8$-matrix $R$ with $R_\ua{}^a=\Ii\, (\gam^a\IC)_{\ua\ub(7)}\,\xi_{2(7)}^\ub$ for $a<8$ and 
$R_\ua{}^8=(k_8)^{-1}\,\IC_{\ua\ub(7)}\,\xi_{2(7)}^\ub$. If $\om^p$ does not depend on $\xi_{1(7)}$ the left hand side of \eqref{8D9} is linear in $\xi_{1(7)}$ and thus \eqref{8D9} implies 
$R_\ua{}^a\, \6\om^p/\6 c^a=0$ which, as $R$ is invertible, implies
$\6\om^p/\6 c^a=0$ for all $c^a$, i.e. $\om^p$ does not depend on translation ghosts at all, which gives $p=0$. This proves the absence of cocycles with $p>0$ which only depend on $\xi_1^-\equiv(0,\Ii\,\xi_{2(7)})$. Analogously one concludes the absence of cocycles with $p>0$ which only depend on $\xi_1^+\equiv (\Ii\,\xi_{1(7)},0)$. As $\fb$ is for $\Ns=1$ homogeneous in both $\xi_1^+$ and $\xi_1^-$
one further infers that for $\Ns=1$ all cocycles with $p>0$ depend at least linearly on both
$\xi_1^+$ and $\xi_1^-$.
In particular this implies for $\Ns=1$ the absence of primitive elements $\vTHE 1{}$ or $\vTHE 1\pm$ as these would be linear in $\xi_1$, as well as the absence of primitive elements $\THE {11}{4}$, $\hTHE {11}{4}$ or $\cTHE {11}\pm{4}$ as these would only involve ghost monomials which do not depend on either $\xi_1^+$ or $\xi_1^-$, see \eqref{dt3} for $\Dim=8$, $p=4$.
 
2. Lemma \ref{lem8D1} differs from the results presented in ref. \cite{Movshev:2011pr} for $\Dim=8$ because there, using the notation of the present paper, cocycles with $\THE{11}3$, $\THE{11,\, a}3$, $\THE{11,\, ab}3$ in place of $\hTHE{11}3$, $\hTHE{11,\, a}3$, $\hTHE{11,\, ab}3$ are presented. This difference is essential because $\THE{11}3$, $\THE{11,\, a}3$, $\THE{11,\, ab}3$ vanish in $\Dim=8$ for our choice of $\CC$ as the matrices $\gam_{abc}\IC$ are antisymmetric (see \eqref{dt1} for $\Dim=8$, $p=3$). An alternative choice of $\CC$ for which the matrices $\gam_a\IC$ are antisymmetric would not resolve this discrepancy because that choice would forbid $\Ns=1$ \cite{note0}.

%Addition:
3. Lemma \ref{lem8D1} does not apply to signatures $(2,6), (6,2)$ because in these cases one has $\Ns\in\{2,4,\dots\}$, see \eqref{i4-3}.

\section{\texorpdfstring{Primitive elements in $\Dim=9$}{Primitive elements in D=9}}\label{9D}

In order to derive the results in $\Dim=9$ by means of the results in $\Dim=8$
we use \eqref{c12} and \eqref{c12a} which give
\begin{align}
a<9:\ &\fb c^a=\tfrac{\Ii}2\,\delta^{ij}\,\xi_{i(8)}\,(\gam^a\IC)_{(8)}\,\xi_{j(8)}^\top
                                =(\fb c^a)_{(8)}\, ,
  \label{9D3}\\
&\fb c^9=\tfrac{\Ii}2\,(k_9)^{-1}\,\delta^{ij}\,\xi_{i(8)}\,(\hat\gam\IC)_{(8)}\,\xi_{j(8)}^\top
  \label{9D3a}
\end{align}
where $(\fb c^a)_{(8)}$ and $\xi_{i(8)}$ denote $\fb c^a$ and $\xi_i$ in $\Dim=8$, with
$\xi_i\equiv \xi_{i(8)}$. 
 
\begin{lemma}[Primitive elements for $\Ns=1$]\label{lem9D1}\quad \\
In the case $\Ns=1$ the general solution of the cocycle condition in $\Hg(\fb)$ is:
\begin{align}
 \fb\om=0\ \LRA\ \om\sim
 \THE{11}4 p_4(\xi)+\THE{11,\, a}4\, p^a_3(\xi)+\THE{11,\, ab}4\, p^{ab}_2(\xi)
 +\THE{11,\, abc}4\, p^{abc}_1(\xi)
 +p_0(\xi)
 \label{9D1} 
\end{align}
with arbitrary polynomials $p_4(\xi)$, $p^a_3(\xi)$, $p^{ab}_2(\xi)$, $p^{abc}_1(\xi)$, $p_0(\xi)$ in the components of $\xi_1$.
\end{lemma}

\PF{\ref{lem9D1}}
Lemma \ref{lem8D1} implies that $\hHg^p(\fb)$ vanishes for $p\geq 4$. Using \eqref{c18} we conclude that $\Hg^p(\fb)$ vanishes for $p> 4$,
\begin{align}
 p>4:\quad \fb\om^p=0\ \LRA\ \om^p\sim 0.
 \label{9D4} 
\end{align}
For $4\geq p\geq 1$ the following implications of \eqref{c12} and \eqref{c12a} are used:
\begin{align}
 &\THE{11}4\propto c^9\,\hTHE{11(8)}3+\ldots\ ,\label{9D6a}
 \\
  a_1,\dots,a_k<9:\ &\THE{11,\, a_1\dots a_k}4=\frac{\6^k \THE{11}4}{\6 c^{a_k}\dots \6 c^{a_1}}\propto c^9\,\hTHE{11,\, a_1\dots a_k(8)}3+\ldots\ ,\label{9D6b}
 \\
 &\THE{11,\, 9\,a_1\dots a_k}4=\frac{\6^{k+1} \THE{11}4}{\6 c^{a_k}\dots \6 c^{a_1}\6 c^9}\propto\hTHE{11,\, a_1\dots a_k(8)}3\label{9D6c}
\end{align}
where ellipses indicate terms without $c^9$.

Using lemma \ref{lem8D1} and exploiting \eqref{c7} in the case $p=1$ \cite{note5} one obtains that one may assume:
\begin{align}
 p=4:\ &\hat\om^3=\hTHE{11(8)}3\, \hat p_4(\xi_{(8)}),\ \hat\om_\mathrm{hom}^4=0;\label{9D5a}
 \\
 p=3:\ &\hat\om^2=\hTHE{11,\, a(8)}3\, \hat p^a_3(\xi_{(8)}),\ \hat\om_\mathrm{hom}^3=\hTHE{11(8)}3\, \hat p^9_3(\xi_{(8)});\label{9D5b}
 \\
 p=2:\ &\hat\om^1=\hTHE{11,\, ab(8)}3\, \hat p^{ab}_2(\xi_{(8)}),\ \hat\om_\mathrm{hom}^2=\hTHE{11,\, a(8)}3\,\hat p^{9\,a}_2(\xi_{(8)});\label{9D5c}
 \\
 p=1:\ &\hat\om^0=\xi_{1(8)}\,(\hat\gam\gam_{abc}\IC)_{(8)}\,\xi_{1(8)}^\top\, \hat p^{abc}_1(\xi_{(8)}),\ 
        \hat\om_\mathrm{hom}^1=\hTHE{11,\, ab(8)}3\, \hat p^{9\, ab}_1(\xi_{(8)}).\label{9D5d}
\end{align}
Using \eqref{9D6a}--\eqref{9D6c} and $\fb\hTHE{11}4=0$ which may be verified explicitly one obtains:
\begin{align}
 \fb\om^4=0\ &\LRA\ \om^4\sim\THE{11}4\, p_4(\xi),\label{9D7a}
 \\
 \fb\om^3=0\ &\LRA\ \om^3\sim\THE{11,\, a}4\, p^a_3(\xi),\label{9D7b}
 \\
 \fb\om^2=0\ &\LRA\ \om^2\sim\THE{11,\, ab}4\, p^{ab}_2(\xi),\label{9D7c}
 \\
 \fb\om^1=0\ &\LRA\ \om^1\sim\THE{11,\, abc}4\, p^{abc}_1(\xi).\label{9D7d}
\end{align}
The lemma is obtained from \eqref{9D4}, \eqref{9D7a}--\eqref{9D7d} and $\om^0=p_0(\xi)$.\QED

The cases $\Ns>1$ can be analyzed analgously to $\Ns>1$ in $\Dim=8$, see proof sketch for lemma \ref{lem8D2}, which gives:

\begin{lemma}[Primitive elements for $\Ns>1$]\label{lem9D2}\quad \\
In the cases $\Ns>1$ the general solution of the cocycle condition in $\Hg(\fb)$ is:
\begin{align}
 \fb\om=0\ \LRA\ \om\sim
 p_0(\xi)
 \label{9D2} 
\end{align}
with an arbitrary polynomial $p_0(\xi)$ in the components of $\xi_1,\ldots,\xi_\Ns$.
\end{lemma}

%Addition:
{\em Comment:} \\
Lemma \ref{lem9D1} does not apply to signatures $(t,9-t)$ with $t\in\{2,3,6,7\}$ because in these cases one has $\Ns\in\{2,4,\dots\}$, see \eqref{i4-3}.

\section{\texorpdfstring{Primitive elements in $\Dim=10$}{Primitive elements in D=10}}\label{10D}

\subsection{Signatures (1,9), (3,7), (5,5), (7,3), (9,1)}\label{10D.1}

In the cases of signatures $(1,9)$, $(3,7)$, $(5,5)$, $(7,3)$, $(9,1)$ the supersymmetry ghosts $\xi_i$ are Majorana Weyl spinors (for signatures $(1,9)$, $(5,5)$, $(9,1)$) or symplectic Majorana Weyl spinors (for signatures $(3,7)$, $(7,3)$). $\Ns_+$ denotes the number of supersymmetry ghosts with positive chirality, $\Ns_-$ denotes the number of supersymmetry ghosts with negative chirality. $\Ns$ is the sum  $\Ns=\Ns_++\Ns_-\in\{1,2,\dots\}$. The case $\Ns=1$ thus includes $(\Ns_+,\Ns_-)=(1,0)$ and $(\Ns_+,\Ns_-)=(0,1)$, the case $\Ns=2$ includes $(\Ns_+,\Ns_-)=(2,0)$, $(\Ns_+,\Ns_-)=(1,1)$ and $(\Ns_+,\Ns_-)=(0,2)$ etc. As in $\Dim=6$ we use $\xi_i=\xi_i^+$ for $i\leq\Ns_+$ and $\xi_i=\xi_i^-$ for $i>\Ns_+$, i.e. the supersymmetry ghosts $\xi_1,\dots,\xi_{\Ns_+}$ have positive chirality and the supersymmetry ghosts $\xi_{\Ns_++1},\dots,\xi_{\Ns_++\Ns_-}$ have negative chirality.

In a spinor representation fulfilling \eqref{c13} and \eqref{c13b} a Weyl spinor $\psi^+=\psi^+\hat\gam$ with positive chirality takes the form $\psi^+=(\chi,0)$ and a Weyl spinor $\psi^-=-\psi^-\hat\gam$ with negative chirality takes the form $\psi^-=(0,\chi)$ where $\chi$ and $0$ have 16 components, respectively, like spinors in $\Dim=9$. In order to derive $\Hg(\fb)$ in $\Dim=10$ by means of $\Hg(\fb)$ in $\Dim=9$, we relate the supersymmetry ghosts $\xi_i$ in $\Dim=10$ to supersymmetry ghosts $\xi_{i(9)}$ in $\Dim=9$ as follows:
\begin{align}
  i\leq \Ns_+:&\ \xi_i\equiv (\xi_{i(9)},0),\quad 
  i>\Ns_+:\ \xi_i\equiv (0,\Ii\,\xi_{i(9)}).\label{10D6}
\end{align}
\eqref{dt4} and \eqref{dt5} for $p=1$ and \eqref{10D6} give:
\begin{align}
a<10:\ &\fb c^a=\ihalf\,\delta^{ij}\,\xi_{i(9)}\,(\gam^a\IC)_{(9)}\,\xi_{j(9)}^\top
                                =(\fb c^a)_{(9)}\, ,
  \label{10D7}
  \\
  &\fb c^{10}=\half\,(k_{10})^{-1}\Big(\sum_{i=1}^{\Ns_+}\xi_{i(9)}\,\IC_{(9)}\,\xi_{i(9)}^\top
  -\sum_{i=\Ns_++1}^{\Ns_++\Ns_-}\xi_{i(9)}\,\IC_{(9)}\,\xi_{i(9)}^\top\Big) .
  \label{10D8}
\end{align}
where $(\fb c^a)_{(9)}$ denotes $\fb c^a$ in $\Dim=9$. Hence, using a spinor representation fulfilling \eqref{c13} and \eqref{c13b} and the identifications \eqref{10D6}, the action of $\fb$ in $\hOmg$ in $\Dim=10$ is identical to the action of $\fb$ in $\Omg$ in $\Dim=9$. This is used to derive $\hHg(\fb)$ in $\Dim=10$ by means of the results for $\Hg(\fb)$ in $\Dim=9$.

\begin{lemma}[Primitive elements for $\Ns_++\Ns_-=1$]\label{lem10D1}\quad \\
In the cases $(\Ns_+,\Ns_-)=(1,0)$ and $(\Ns_+,\Ns_-)=(0,1)$ the general solution of the cocycle condition in $\Hg(\fb)$ is:
\begin{align}
 \fb\om=0\ \LRA\ \om\sim
 &\ \THE{11}5\,p_5(\xi)+\THE{11,\, a}5\,p_4^a(\xi)+\THE{11,\, ab}5\,p_3^{ab}(\xi)\notag
 \\
 &+\THE{11,\,abc}5\,p_2^{abc}(\xi)+
\vTHE{1}{\ua}\, p_\ua(\xi)
 +p_0(\xi)
 \label{10D1} 
\end{align}
with arbitrary polynomials $p_0(\xi)$, $p_\ua(\xi)$, $p_2^{abc}(\xi)$, $p_3^{ab}(\xi)$,
$p_4^a(\xi)$, $p_5(\xi)$ in the components of $\xi_1$.
\end{lemma}

\PF{\ref{lem10D1}}
Lemma \ref{lem9D1} implies that $\hHg^p(\fb)$ vanishes for $p\geq 5$. Using \eqref{c18} we conclude that $\Hg^p(\fb)$ vanishes for $p> 5$,
\begin{align}
 p>5:\quad \fb\om^p=0\ \LRA\ \om^p\sim 0.
 \label{10D9} 
\end{align}
For $5\geq p\geq 1$ the following implications of \eqref{c13} and \eqref{c13b} are used:
\begin{align}
 &\THE{11}5\propto c^{10}\,\THE{11(9)}4+\ldots\ ,\label{10D6a}
 \\
  a_1,\dots,a_k<10:\ &\THE{11,\, a_1\dots a_k}5=\frac{\6^k \THE{11}5}{\6 c^{a_k}\dots \6 c^{a_1}}\propto c^{10}\,\THE{11,\, a_1\dots a_k(9)}4+\ldots\ ,\label{10D6b}
 \\
 &\THE{11,\, 10\,a_1\dots a_k}5=\frac{\6^{k+1} \THE{11}5}{\6 c^{a_k}\dots \6 c^{a_1}\6 c^{10}}\propto\THE{11,\, a_1\dots a_k(9)}4\label{10D6c}
\end{align}
where ellipses indicate terms without $c^{10}$.

Using lemma \ref{lem9D1} and exploiting \eqref{c7} in the case $p=1$ \cite{note6} one obtains that one may assume:
\begin{align}
 p=5:\ &\hat\om^4=\THE{11(9)}4\, \hat p_5(\xi_{(9)}),\ \hat\om_\mathrm{hom}^5=0;\label{10D5a}
 \\
 p=4:\ &\hat\om^3=\THE{11,\, a(9)}4\, \hat p^a_4(\xi_{(9)}),\ 
       \hat\om_\mathrm{hom}^4=\THE{11(9)}4\,\hat p^{10}_4(\xi_{(9)});\label{10D5b}
 \\
 p=3:\ &\hat\om^2=\THE{11,\, ab(9)}4\, \hat p^{ab}_3(\xi_{(9)}),\ 
       \hat\om_\mathrm{hom}^3=\THE{11,\, a(9)}4\,\hat p^{10\, a}_4(\xi_{(9)});\label{10D5c}
 \\
 p=2:\ &\hat\om^1=\THE{11,\, abc(9)}4\, \hat p^{abc}_2(\xi_{(9)}),\ 
        \hat\om_\mathrm{hom}^2=\THE{11,\, ab(9)}4\,\hat p^{10\, ab}_2(\xi_{(9)});\label{10D5d}
 \\
 p=1:\ &\hat\om^0=\xi_{1(9)}^\ua\,\hat p_\ua (\xi_{(9)}) ,\ 
        \hat\om_\mathrm{hom}^1=\THE{11,\, abc(9)}4\, \hat p^{10\,abc}_2(\xi_{(9)}).\label{10D5e}
\end{align}
Using \eqref{10D6a}--\eqref{10D6c} as well as $\fb\hTHE{11}5=0$ and $\fb\vTHE{1}{\ua}=0$ which may be verified explicitly, one obtains \cite{note6}:
\begin{align}
 \fb\om^5=0\ &\LRA\ \om^5\sim\THE{11}5\,p_5(\xi),\label{10D7a}
 \\
 \fb\om^4=0\ &\LRA\ \om^4\sim\THE{11,\, a}5\,p_4^{a}(\xi),\label{10D7b}
 \\
 \fb\om^3=0\ &\LRA\ \om^3\sim\THE{11,\, ab}5\,p_3^{ab}(\xi),\label{10D7c}
 \\
 \fb\om^2=0\ &\LRA\ \om^2\sim\THE{11,\,abc}5\,p_2^{abc}(\xi),\label{10D7d}
 \\
 \fb\om^1=0\ &\LRA\ \om^1\sim\vTHE{1}{\ua}\, p_\ua(\xi).\label{10D7e}
\end{align}
The lemma is obtained from \eqref{10D9}, \eqref{10D7a}--\eqref{10D7e} and $\om^0=p_0(\xi)$.\QED

The results for the cases $\Ns_++\Ns_-\geq 2$ can be derived by means of lemmas \ref{lem9D1}, \ref{lem9D2} and \ref{lem10D1} analogously to the derivation of lemmas \ref{lem6D2} and \ref{lem6D3} by means of lemmas \ref{lem5D1}, \ref{lem5D2} and \ref{lem6D1}. One obtains:

\begin{lemma}[Primitive elements for $\Ns_++\Ns_-=2$]\label{lem10D2}\quad \\
(i) In the case $(\Ns_+,\Ns_-)=(1,1)$ the general solution of the cocycle condition in $\Hg(\fb)$ is:
\begin{align}
 \fb\om=0\ \LRA\ \om\sim
 (\THE{11}1-\THE{22}1)\,p_1(\xi)+p_0(\xi)
 \label{10D2a} 
\end{align}
with arbitrary polynomials $p_0(\xi)$, $p_1(\xi)$ in the components of $\xi_1,\xi_2$.

(ii) In the cases $(\Ns_+,\Ns_-)=(2,0)$ and $(\Ns_+,\Ns_-)=(0,2)$ the general solution of the cocycle condition in $\Hg(\fb)$ is:
\begin{align}
 \fb\om=0\ \LRA\ \om\sim
 \THE{12}1\, p^{12}_1(\xi)+
 (\THE{11}1-\THE{22}1)\,p_1(\xi)+p_0(\xi)
 \label{10D2} 
\end{align}
with arbitrary polynomials $p_0(\xi)$, $p_1(\xi)$, $p^{12}_1(\xi)$ in the components of $\xi_1,\xi_2$.
\end{lemma}

\begin{lemma}[Primitive elements for $\Ns_++\Ns_->2$]\label{lem10D3}\quad \\
In the cases $\Ns_++\Ns_->2$ the general solution of the cocycle condition in $\Hg(\fb)$ is:
\begin{align}
 \fb\om=0\ \LRA\ \om\sim
 p_0(\xi)
 \label{10D3} 
\end{align}
with an arbitrary polynomial $p_0(\xi)$ in the components of $\xi_1,\dots,\xi_\Ns$.
\end{lemma}

{\em Comments:} \\
1. The difference between the results for $(\Ns_+,\Ns_-)=(1,1)$ and for $(\Ns_+,\Ns_-)\in\{(2,0),$ $(0,2)\}$ parallels the situation in $\Dim=2$ \cite{Brandt:2010fa} and $\Dim=6$.

%Addition:
2. We note that in lemma \ref{lem10D1} one has $\THE{11}5=\cTHE{11}+5$ and $\vTHE{i}{}=\vTHE{1}-$ in the case $(\Ns_+,\Ns_-)=(1,0)$, and $\THE{11}5=\cTHE{11}-5$ and $\vTHE{1}{}=\vTHE{1}+$ in the case $(\Ns_+,\Ns_-)=(0,1)$.
Analogously in lemma \ref{lem10D2} one has $\THE{ij}1=\cTHE{ij}+1$ in the case $(\Ns_+,\Ns_-)=(2,0)$, and $\THE{ij}1=\cTHE{ij}-1$ in the case $(\Ns_+,\Ns_-)=(0,2)$.

3. Lemma \ref{lem10D1} only applies to signatures $(1,9), (5,5), (9,1)$ because
in the cases of signatures $(3,7), (7,3)$ one has $\Ns\in\{2,4,\dots\}$, see \eqref{i4-3}.

\subsection{Signatures (0,10), (2,8), (4,6), (6,4), (8,2), (10,0)}\label{10D.2}

In the cases of signatures $(0,10)$, $(2,8)$, $(4,6)$, $(6,4)$, $(8,2)$, $(10,0)$ the supersymmetry ghosts $\xi_i$ are Majorana spinors consisting of two Weyl spinors with opposite chiralities, respectively. $\Ns\in\{1,2,\dots\}$ denotes the number of these Majorana supersymmetry ghosts. Hence, there are both $\Ns$ Weyl supersymmetry ghosts with positive chirality and $\Ns$ Weyl supersymmetry ghosts with negative chirality. 

The case $\Ns=1$ corresponds thus to the case $(\Ns_+,\Ns_-)=(1,1)$ in lemma \ref{lem10D2}. Using $\xi_1\,\hat\gam=\xi^+_1-\xi^-_1$ and identifying $\xi^+_1,\xi^-_1$ with $\xi_1,\xi_2$ in lemma \ref{lem10D2}, respectively, lemma \ref{lem10D2} gives directly:

\begin{lemma}[Primitive elements for $\Ns=1$]\label{lem10D4}\quad \\
In the case $\Ns=1$ the general solution of the cocycle condition in $\Hg(\fb)$ is:
\begin{align}
 \fb\om=0\ \LRA\ \om\sim
 \hTHE{11}1\, p_1(\xi)+p_0(\xi)
 \label{10D4} 
\end{align}
with arbitrary polynomials $p_0(\xi)$, $p_1(\xi)$ in the components of $\xi_1$.
\end{lemma}

The cases $\Ns>1$ correspond to cases $\Ns_+=\Ns_->1$ in lemma \ref{lem10D3} which implies: 

\begin{lemma}[Primitive elements for $\Ns>1$]\label{lem10D5}\quad \\
In the cases $\Ns>1$ the general solution of the cocycle condition in $\Hg(\fb)$ is:
\begin{align}
 \fb\om=0\ \LRA\ \om\sim
 p_0(\xi)
 \label{10D5} 
\end{align}
with an arbitrary polynomial $p_0(\xi)$ in the components of $\xi_1,\dots,\xi_\Ns$.
\end{lemma}

\section{\texorpdfstring{Primitive elements in $\Dim=11$}{Primitive elements in D=11}}\label{11D}

The results in $\Dim=11$ are derived by means of the results in $\Dim=10$
presented in section \ref{10D.2}. To this end we use \eqref{c12} and \eqref{c12a} which give
\begin{align}
a<11:\ &\fb c^a=\tfrac{\Ii}2\,\delta^{ij}\,\xi_{i(10)}\,(\gam^a\IC)_{(10)}\,\xi_{j(10)}^\top
                                =(\fb c^a)_{(10)}\, ,
  \label{11D3}\\
&\fb c^{11}=\tfrac{\Ii}2\,(k_{11})^{-1}\,\delta^{ij}\,\xi_{i(10)}\,(\hat\gam\IC)_{(10)}\,\xi_{j(10)}^\top
  \label{11D3a}
\end{align}
where $(\fb c^a)_{(10)}$ and $\xi_{i(10)}$ denote $\fb c^a$ and $\xi_i$ in $\Dim=10$, each $\xi_{i(10)}$ being a 32-component spinor consisting of two Weyl spinors, with 
$\xi_i\equiv \xi_{i(10)}$. \eqref{11D3} shows that we can 
use lemmas \ref{lem10D4} and \ref{lem10D5} to obtain $\hHg(\fb)$ in $\Dim=11$.
 
\begin{lemma}[Primitive elements for $\Ns=1$]\label{lem11D1}\quad \\
In the case $\Ns=1$ the general solution of the cocycle condition in $\Hg(\fb)$ is:
\begin{align}
 \fb\om=0\ \LRA\ \om\sim
 \THE{11}2\, p_2(\xi)+\THE{11,\, a}2\, p^a_1(\xi)
 +p_0(\xi)
 \label{11D1} 
\end{align}
with arbitrary polynomials $p_0(\xi)$, $p^a_1(\xi)$, $p_2(\xi)$ in the components of $\xi_1$.
\end{lemma}

\PF{\ref{lem11D1}}
Lemma \ref{lem10D4} implies that $\hHg^p(\fb)$ vanishes for $p\geq 2$. Using \eqref{c18} we conclude that $\Hg^p(\fb)$ vanishes for $p> 2$,
\begin{align}
 p>2:\quad \fb\om^p=0\ \LRA\ \om^p\sim 0.
 \label{11D6} 
\end{align}
For $p=2$ and $p=1$ the following implications of \eqref{c12} and \eqref{c12a} are used:
\begin{align}
 &\THE{11}2\propto c^{11}\,\hTHE{11(10)}1+\ldots\ ,\label{11D8}
 \\
  a<11:\ &\THE{11,\, a}2=\frac{\6 \THE{11}2}{\6 c^a}\propto c^{11}\,\frac{\6 \hTHE{11(10)}1}{\6 c^a}+\ldots=
 c^{11}\,\xi_{1(10)}\,(\hat\gam\gam_a\IC)_{(10)}\, \xi_{1(10)}^\top+\ldots\ ,\label{11D8a}
 \\
 &\THE{11,\, 11}2=\frac{\6 \THE{11}2}{\6 c^{11}}\propto\hTHE{11(10)}1
 \label{11D8b} 
\end{align}
where ellipses indicate terms without $c^{11}$. 

Using lemma \ref{lem10D4} and exploiting \eqref{c7} in the case $p=1$ analogously as in $\Dim=9$ \cite{note5} one obtains that one may assume:
\begin{align}
 p=2:\ &\hat\om^1=\hTHE{11(10)}1\, \hat p_2(\xi_{(10)}),\quad \hat\om^2_{\mathrm{hom}}=0;
 \label{11D7} \\
 p=1:\ &\hat\om^0=\xi_{1(10)}\,(\hat\gam\gam_a\IC)_{(10)}\, \xi_{1(10)}^\top\,\hat p^a_1(\xi_{(10)}),\ \hat\om^1_{\mathrm{hom}}=\hTHE{11(10)}1\, \hat p^{11}_1(\xi_{(10)}).\label{11D10}
\end{align}
Using \eqref{11D8}--\eqref{11D8b} and $\fb\THE{11}2=0$ which may be verified explicitly one obtains:
\begin{align}
 \fb\om^2=0\ \LRA\ &\om^2\sim \THE{11}2\, p_2(\xi),\label{11D9} 
 \\
 \fb\om^1=0\ \LRA\ &\om^1\sim \THE{11,\, a}2\, p^a_1(\xi).
 \label{11D12}
\end{align} 
The lemma is obtained from \eqref{11D6}, \eqref{11D9}, \eqref{11D12} and $\om^0=p_0(\xi)$. \QED

The cases $\Ns>1$ can be analyzed analogously to $\Ns>1$ in $\Dim=8$, see proof sketch for lemma \ref{lem8D2}, which gives:

\begin{lemma}[Primitive elements for $\Ns>1$]\label{lem11D2}\quad \\
In the cases $\Ns>1$ the general solution of the cocycle condition in $\Hg(\fb)$ is:
\begin{align}
 \fb\om=0\ \LRA\ \om\sim p_0(\xi)
 \label{11D2} 
\end{align}
with an arbitrary polynomial $p_0(\xi)$ in the components of $\xi_1,\dots,\xi_\Ns$.
\end{lemma}

%Addition:
{\em Comment:} \\
Lemma \ref{lem11D1} does not apply to signatures $(t,11-t)$ with $t\in\{0,3,4,7,8,11\}$ because in these cases one has $\Ns\in\{2,4,\dots\}$, see \eqref{i4-3}.

\renewcommand\refname{ }%entfernt �berschrift "References"


\begin{thebibliography}{99}
\bibitem{note1}
The first equation \eqref{5D5} and lemma \ref{lem5D1} imply $\om^1_{\ol{m}}=\sum_{i,j\leq 2}\THE{ij,\, a}2 p^{ij\,a}(\xi)+\fb^{(1)}\eta^2$ for some polynomials $p^{ij\,a}(\xi)$ and some $\eta^2\in\Omg^2$. $\fb^{(2)}\THE{11,\, a}2$ does not depend on $\xi_2$, $\fb^{(2)}\THE{22,\, a}2$ does not depend on $\xi_1$ and $\fb^{(2)}\THE{12,\, a}2=\fb^{(2)}\THE{21,\, a}2$ is symmetric under $\xi_1\leftrightarrow\xi_2$ whereas all terms in $\fb^{(1)}$ depend on both $\xi_1$ and $\xi_2$ and $\fb^{(1)}$ is antisymmetric under $\xi_1\leftrightarrow\xi_2$.
As a consequence the second equation \eqref{5D5} implies that the $p^{11\,a}(\xi)$, $p^{12\,a}(\xi)$, $p^{22\,a}(\xi)$ are such that $\om^1_{\ol{m}}$ is $\fb^{(1)}$-exact (or vanishes).
\bibitem{note2a}
In $\Dim=6$ one has for $(\Ns_+,\Ns_-)=(2,0)$ and $(\Ns_+,\Ns_-)=(0,2)$:
\begin{align}
 \fb\THE {ij}3&=
 \ihalf\,(\xi_1\,\gam^a\IC\,\xi_2^\top)\,(\xi_i\,\gam_{abc}\IC\,\xi_j^\top)\,c^b c^c
 \notag\\ 
 &=\sum_{a\not\in \{b,c\}}\ihalf\,
 (\xi_1\,\gam^a\IC\,\xi_2^\top)\,(\xi_{i}^{}\,\gam_a\gam_b\gam_c\IC\,\xi_{j}^\top)\,c^b c^c\notag\\
 &=\ihalf\,(\xi_1\,\gam^a\IC\,\xi_2^\top)\,(\xi_{(i}^{}\,\gam_a\gam_b\gam_c\IC\,\xi_{j)}^\top)\,c^b c^c\notag\\
 &\phantom{=}-\ihalf\,(\xi_1\,\gam^b\IC\,\xi_2^\top)\,(\xi_{(i}^{}\,\gam_b\gam_b\gam_c\IC\,\xi_{j)}^\top)\,c^b c^c\notag\\
 &\phantom{=}-\ihalf\,(\xi_1\,\gam^c\IC\,\xi_2^\top)\,(\xi_{(i}^{}\,\gam_c\gam_b\gam_c\IC\,\xi_{j)}^\top)\,c^b c^c\notag\\
 &=\half\,(\fb\vTHE{(i}{})\,\gam_b\gam_c\IC\,\xi_{j)}^\top\,c^b c^c
 -\Ii\,(\xi_1\,\gam_b\IC\,\xi_2^\top)\,(\xi_{(i}^{}\,\gam_c\IC\,\xi_{j)}^\top)\,c^b c^c=0\notag
\end{align}
where we used that $\fb\vTHE{i}{}$ vanishes for $(\Ns_+,\Ns_-)=(2,0)$ and $(\Ns_+,\Ns_-)=(0,2)$ \cite{note2b}, and that
$\xi_{(i}^{}\,\gam_c\IC\,\xi_{j)}^\top$ also vanishes because $\gam_c\IC$ is antisymmetric in $\Dim=6$, see \eqref{dt1}.
\bibitem{note2b}
\eqref{c13} and \eqref{6D6} imply $\vTHE i{}=(0,-\Ii\,k_6\,c^6\,\xi_{i(5)}+\vTHE {i(5)}{})$ for $(\Ns_+,\Ns_-)=(2,0)$, 
and
$\vTHE i{}=(-k_6\,c^6\,\xi_{i(5)}+\Ii\,\vTHE {i(5)}{},0)$ for $(\Ns_+,\Ns_-)=(0,2)$. $\fb\vTHE i{}=0$ follows from the ``completeness relation''
of the $\gam$-matrices in $\Dim=5$ which reads 
\begin{align}
\Dim=5:\ \delta_\ua^\ub\delta_\ug^\ud+\gam^a{}_\ua{}^\ub\gam_{a\,\ug}{}^\ud
-\gam^{ab}{}_\ua{}^\ub\gam_{ab\,\ug}{}^\ud=4\delta_\ua^\ud\delta_\ug^\ub
\notag
\end{align}
and implies 
\begin{align}
(\fb\vTHE {i}\ua)_{(5)}=
\Ii\,\xi_{1(5)}(\gam^a\IC)_{(5)}\xi_{2(5)}^\top\,
(\xi_{i(5)}\gam_{a(5)})^\ua=
\Ii\,\xi_{1(5)}\IC_{(5)}\xi_{2(5)}^\top\,\xi_{i(5)}^\ua
\notag
\end{align}
where we used that $(\gam^a\IC)_{(5)}$ and $\IC_{(5)}$ are antisymmetric, whereas $(\gam^{ab}\IC)_{(5)}$ is symmetric, see \eqref{dt1}.
\bibitem{note2}
For instance:
\begin{align}
 \THE{ij,\, 1(5)}2&\propto \frac{\6^2\THE{ij}3}{\6 c^1\6 c^6}=
 \sum_{a=2}^5 c^a\xi_i\,\gam_a\gam_6\gam_1\IC\,\xi_j^\top
 \notag\\
&=\half\,(\vTHE i{}-c^1\xi_i\,\gam_1-c^6\xi_i\,\gam_6)\,\gam_6\gam_1\IC\,\xi_j^\top
+(i\leftrightarrow j)
\notag\\
 &=\half\,\vTHE i{}\,\gam_6\gam_1\IC\,\xi_j^\top+(i\leftrightarrow j)
 =\vTHE {(i}{}\,\gam_{61}\IC\,\xi_{j)}^\top\notag
\end{align}
where we used that $c^1\xi_{(i}{}\,\gam_{6}\IC\,\xi_{j)}^\top$ and $c^6\xi_{(i}{}\,\gam_{1}\IC\,\xi_{j)}^\top$
vanish because $\gam_a\IC$ is antisymmetric in $\Dim=6$, see \eqref{dt1}.
\bibitem{note3}
Using the structure of $\fb c^7$ (which consists of bilinears $\xi_{1(6)}^+\xi_{2(6)}^-$ and $\xi_{1(6)}^-\xi_{2(6)}^+$) and $\fb c^1, \dots,\fb c^6$ (which only involve bilinears $\xi_{1(6)}^+\xi_{2(6)}^+$ and $\xi_{1(6)}^-\xi_{2(6)}^-$) one infers from \eqref{c7} for $p=1$ that $\hat\om^0$ is at least linear in both $\xi_{1(6)}$ and $\xi_{2(6)}$. This gives $\hat\om^0=\xi_{1(6)}^\ua\xi_{2(6)}^\ub
K_{\ua\ub}^r\,p_r(\xi_{(6)})$ with constant $8\times 8$-matrices $K^r$ which can be chosen according to   
$K^r\in\{(\IC)_{(6)},$ 
$(\gam_a\IC)_{(6)},$
$(\gam_{ab}\IC)_{(6)},$
$(\gam_{abc}\IC)_{(6)},$
$(\hat\gam\IC)_{(6)},$
$(\hat\gam\gam_{a}\IC)_{(6)},$
$(\hat\gam\gam_{ab}\IC)_{(6)}\}$ since these matrices form a basis for $8\times 8$-matrices.
\eqref{c7} then leads to $\hat\om^0$ as in \eqref{7D10}, up to a coboundary in $\hHg(\fb)$ which can be neglected.
\bibitem{note4}
For $\Ns=1$ all cocycles with $p>0$ depend at least linearly on both
$\xi_{1(7)}$ and $\xi_{2(7)}$, see first comment at the end of section \ref{8D}.
In the case $p=1$ this implies $\hat\om^0=\xi_{1(7)}^\ua\xi_{2(7)}^\ub
K_{\ua\ub}^r\,p_r(\xi_{(7)})$ with constant $8\times 8$-matrices $K^r$ which can be chosen according to   
$K^r\in\{(\IC)_{(7)},$ 
$(\gam_a\IC)_{(7)},$
$(\gam_{ab}\IC)_{(7)},$
$(\gam_{abc}\IC)_{(7)}\}$ since these matrices form a basis for $8\times 8$-matrices. 
\eqref{c7} then leads to $\hat\om^0$ as in \eqref{8D5c}, up to a coboundary in $\hHg(\fb)$ which can be neglected.
\bibitem{note5}
The structures of $\fb c^9$ (which consists of bilinears $\xi_{1(8)}^+\xi_{1(8)}^+$ and $\xi_{1(8)}^-\xi_{1(8)}^-$) and $\fb c^1, \dots,\fb c^8$ (which only involve bilinears $\xi_{1(8)}^+\xi_{1(8)}^-$) imply that
$\hat\om^0=\xi_{1(8)}^\ua\xi_{1(8)}^\ub
K_{\ua\ub}^r\,p_r(\xi_{(8)})$ with constant $16\times 16$-matrices $K^r$ which can be chosen according to   
$K^r\in\{(\IC)_{(8)},$ 
$(\gam_a\IC)_{(8)},$ \dots, $(\gam_{abcd}\IC)_{(8)}$,
$(\hat\gam\IC)_{(8)}$,
$(\hat\gam\gam_a\IC)_{(8)},$ \dots, $(\hat\gam\gam_{abc}\IC)_{(8)}
\}$ since these matrices form a basis for $16\times 16$-matrices. 
\eqref{c7} then leads to $\hat\om^0$ as in \eqref{9D5d}, up to a coboundary in $\hHg(\fb)$ which can be neglected.
\bibitem{note6}
In the case $p=1$ we use that every cocycle $\om^1$ must be at least linear in the supersymmetry ghosts since no nonvanishing linear combination of the $c^a$ with constant coefficients is $\fb$-closed. This yields $\hat\om^0$ as in \eqref{10D5e}. Lemma \ref{lem9D1} yields $\hat \om^1_\mathrm{hom}$ as in \eqref{10D5e}. \eqref{10D7e} follows from the facts that every monomial $c^{10}\xi_{1(9)}^\ua$ corresponds to a cocycle in $\Hg^1(\fb)$ proportional to one of the $\vTHE{1}{\ua}$, and that $\THE{11,\, abc(9)}4\propto \vTHE 1{}\gam_{10 abc}\IC\xi_1^\top$, analogously to the situation in $\Dim=6$ \cite{note2b,note2}.
\end{thebibliography}
\end{document}